\documentclass[aps,twocolumn,showpacs,preprintnumbers,nofootinbib,prd,superscriptaddress,groupedaddress,10pt]{revtex4-2}

\makeatletter
\def\l@subsubsection#1#2{}
\def\l@subsubsubsection#1#2{}
\makeatother

\usepackage{graphicx,amssymb,amsmath,amsthm,amsfonts,epsfig,epsf,fixmath}
\usepackage[usenames]{color}
\usepackage{epstopdf}

\usepackage{bm}
\usepackage{amsmath}
\usepackage{dcolumn}
 \usepackage[latin1]{inputenc}
\usepackage{latexsym}
\usepackage{rotating}
\usepackage{longtable}

\usepackage{enumerate}
\usepackage{tensor,multirow}
\usepackage{mathtools}
\usepackage{url}
\usepackage[linktocpage]{hyperref}

\begin{document}
\title{Black holes immersed in dark matter: energy condition and sound speed}

\newcommand{\MPI}{Max-Planck-Institut f{\"u}r Gravitationsphysik (Albert-Einstein-Institut), D-30167 Hannover, Germany}

\newcommand{\LBNZ}{Leibniz Universit{\"a}t Hannover, D-30167 Hannover, Germany}

\author{Sayak Datta}\email{sayak.datta@aei.mpg.de}
\affiliation{\MPI}\affiliation{\LBNZ}

\date{\today}

\begin{abstract}
   In this work, we study the impact of the environment around a black hole in detail. We introduce non-vanishing radial pressure in a manner analogous to compact stars. We examine both isotropic and anisotropic fluid configurations with and without radial pressure respectively. Our focus extends beyond just dark matter density to the vital role of the energy condition and sound speed in the spacetime of a black hole immersed in matter. In cases of anisotropic pressure with vanishing radial pressure, all profiles violate the dominant energy condition near the BH, and the tangential sound speed exceeds light speed for all dark matter profiles. In our second approach, without assuming vanishing radial pressure, we observe similar violations and superluminal sound speeds. To rectify this, we introduce a hard cutoff for the sound speed, ensuring it remains subluminal. As a consequence, the energy condition is also satisfied. However, this results in increased density and pressure near the BH. This raises questions about the sound speed and its impact on the density structure, as well as questions about the validity of the model itself. With the matter distribution, we also compute the metric for different configurations. It reveals sensitivity to the profile structure. The metric components point towards the horizon structure. 
\end{abstract}

\maketitle

\section{Introduction}\label{sec:intro}

Substantial evidence firmly supports the presence of dark matter within the central regions of the galaxies \cite{Freese:2008cz, Navarro:1995iw, Clowe:2006eq, Bertone:2004pz}. Apart from its established gravitational interaction, the exact nature of dark matter remains enigmatic. Unraveling the properties of dark matter within the framework of the ``Standard Model" of particle physics remains a significant scientific endeavor. Despite persistent efforts to identify potential, albeit minuscule, interactions between the dark matter and the standard model, these pursuits have thus far not yielded definitive results. Nonetheless, the quest to detect these interactions will persist with great vigor in the years ahead \cite{Kahlhoefer:2017dnp, PerezdelosHeros:2020qyt}.

Alongside, probing the dynamics of the accreting baryonic matter in the galactic centers is crucial for astrophysics. Astrophysical compact sources, including binary systems, do not exist in isolation but evolve within complex environments comprising plasma, electromagnetic fields, along with the dark matter (DM) \cite{Yunes:2011ws, Barausse:2014tra, Cardoso:2019rou, Cardoso:2020iji, Derdzinski:2020wlw, Cardoso:2021wlq, Zwick:2022dih}. 
With the advent of gravitational wave (GW) astronomy  \cite{LIGOScientific:2016aoc, LIGOScientific:2020ibl} and very-long baseline interferometry \cite{EventHorizonTelescope:2019dse, GRAVITY:2020gka}, the era of the multimessenger astronomy revolution has already begun. These observatories are specifically designed to investigate compact objects, including black holes (BHs). In time it will revolutionize our ability to study the invisible universe \cite{Barack:2018yly, Cardoso:2019rvt, Bertone:2018krk, Bar:2019pnz, Brito:2015oca}. In light of these advancements, we can ask how the dark and baryonic matter around an inspiraling binary impacts the binary evolution.

Dark matter, which may cluster at the center of galaxies and close to BHs \cite{Gondolo:1999ef, Sadeghian:2013laa}, could significantly impact the dynamics of compact binaries and how GWs or electromagnetic waves propagate \cite{Barack:2018yly, Eda:2013gg, Macedo:2013qea, Barausse:2014tra, Baibhav:2019rsa, Seoane:2021kkk}. This prompts the question of how the existence of surrounding matter impacts the generation and transmission of GWs, as well as the electromagnetic characteristics of BHs. To answer this question, the knowledge of motivated matter distributions and corresponding spacetime geometry will be extremely useful.

Gaining insights into how the distribution of matter influences the behavior of merging binary systems and discerning the resulting impact on GW production and propagation mechanisms necessitates comprehensive relativistic solutions that describe BHs within a medium. These environmental effects offer a new route to determine fundamental astrophysical properties, such as the distribution of dark and baryonic matter surrounding massive objects, as well as shedding light on accretion phenomena \cite{DeLuca:2022xlz, Sberna:2022qbn, Speri:2022upm, Kavanagh:2020cfn, Speeney:2022ryg, Macedo:2013qea, Eda:2013gg}.

Currently, existing research has predominantly gravitated toward Newtonian methodologies. These approaches often rely on the slow-motion quadrupole formula for estimating GW emission and dynamics \cite{Babak:2006uv, Destounis:2021mqv}, or they focus on Newtonian dynamical friction \cite{Speeney:2022ryg, Vicente:2022ivh, Traykova:2021dua, Sadeghian:2013laa}. While certain studies have ventured into modeling effects like gravitational redshift or peculiar motion by invoking Doppler-like waveform adjustments \cite{Tamanini:2019usx}, a first-principle derivation is still required.

Nevertheless, recent efforts have emerged to enhance the analysis by incorporating certain relativistic effects, highlighting their potential to substantially influence the conclusions concerning detectability and parameter estimation. Ref. \cite{ Cardoso:2022whc} has introduced spacetime geometry generated by a non-spinning BH within a core of matter, offering a first step towards studying BH physics in realistic dense environments for a specific choice of matter distribution. In Ref. \cite{Konoplya:2022hbl, Figueiredo:2023gas, Igata:2022rcm} generic matter profiles were considered. They explore the influence of various families of DM halos on the geometry of BHs and their associated geodesic structures. Recently it has also been extended to wormholes \cite{Biswas:2023ofz}. These profiles have been used to compute the impact of DM profiles on the GW fluxes from extreme mass ratio inspirals immersed in such environment \cite{Cardoso:2022whc, Cardoso:2022whc, Figueiredo:2023gas, Rahman:2023sof}

Nonetheless, several aspects remain to be explored. Firstly, previous works assumed the radial pressure to be negligible. In our present study, we take into account a non-zero radial pressure. Secondly, it becomes evident that the matter structures examined in prior works and our current work, exhibit unphysical sound speeds and, in specific instances, diverging anomalies. We address and mitigate some of these issues, examining their effects on the matter distribution and as a consequence, on the geometry.

\begin{figure*}\includegraphics[width=85mm]{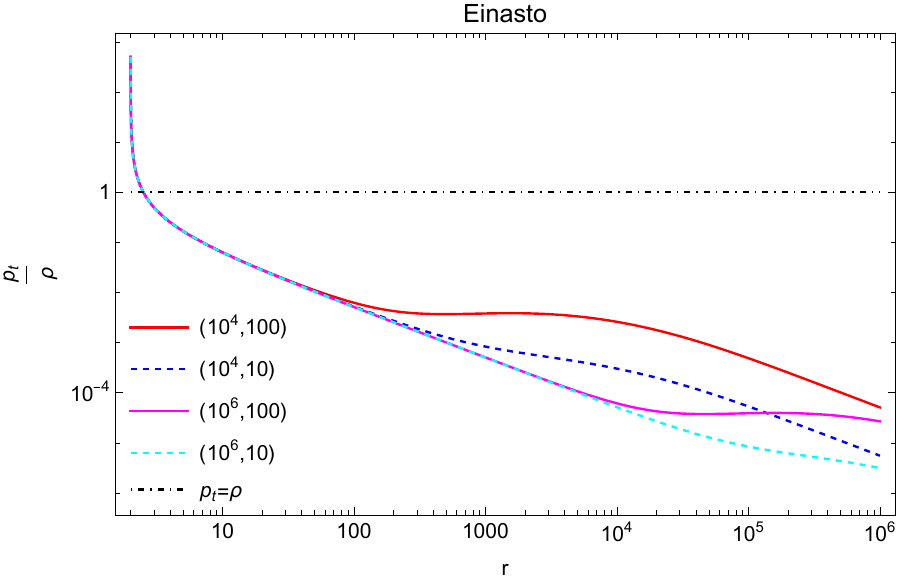}
\includegraphics[width=85mm]{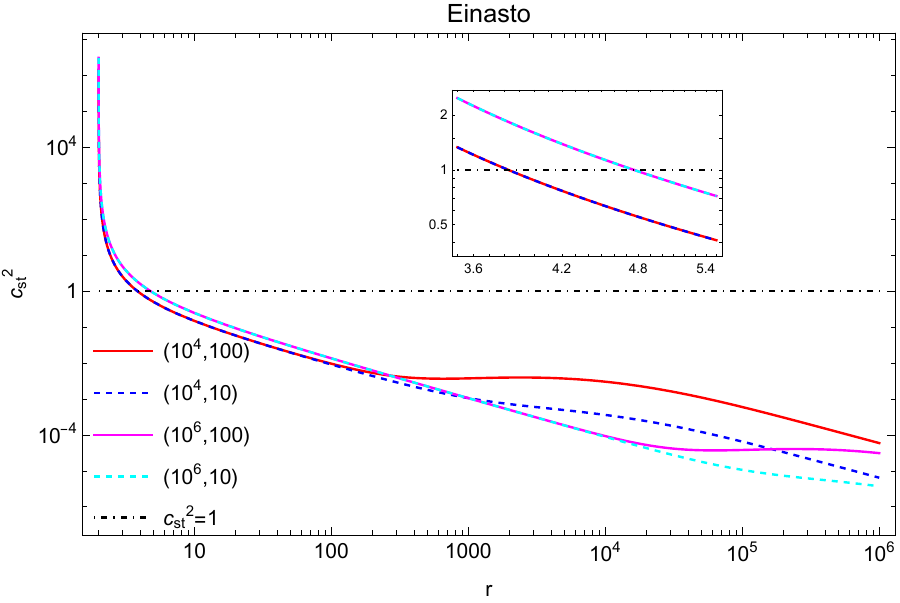}
\includegraphics[width=85mm]{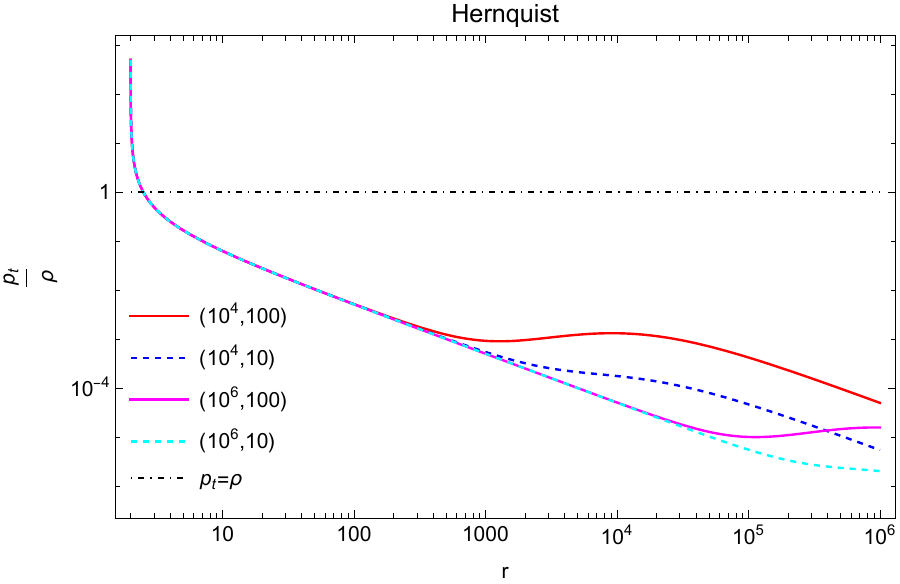}
\includegraphics[width=85mm]{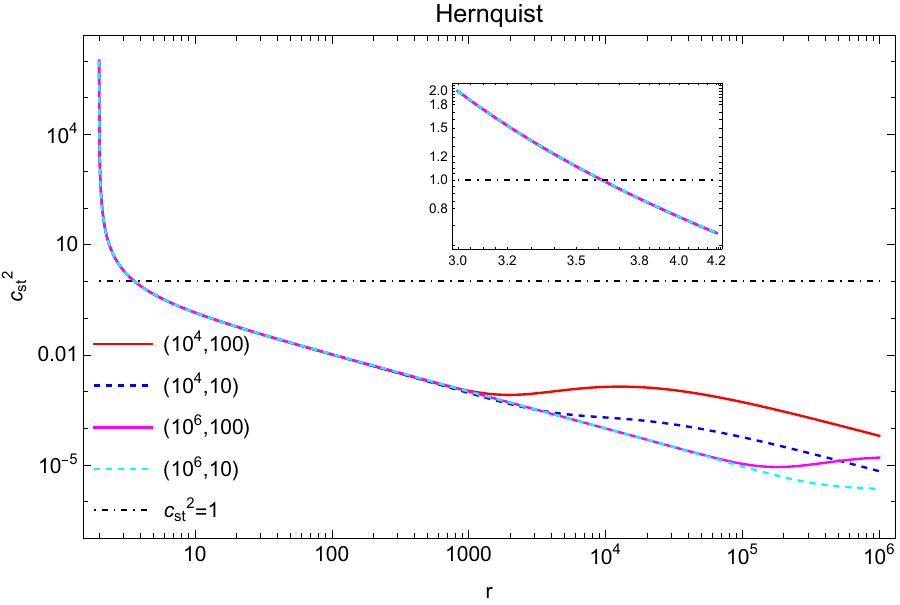}
\caption{In this figure we consider $p_r=0$. In the left column, we show the behavior of the dominant energy condition with respect to the radius, down to $r=2.001$, for both the Einasto and Hernquist profiles. The condition is violated near the BH. In the right column, we show the behavior of tangential sound speed with respect to the radius for both the Einasto and Hernquist profiles. Near the BH the sound speed becomes faster than light. In the subplot, the region where the sound speed becomes unity is demonstrated.}
\label{fig:tangential cs}
\end{figure*}

In Section \ref{sec:General structure for individual distribution}, we delve into the structural aspects of the matter profiles under consideration. Section \ref{sec:Sound speed problem} provides a comprehensive analysis of the pressure equation solutions, focusing on issues associated with sound speed and the measures taken to address these problems. The results of these mitigations are presented. Moving forward, Section \ref{sec:Metric solutions with DM profile} explores the computation of the metric using the derived density and pressure profiles, shedding light on its structure in the vicinity of the black hole (BH) horizon. In Section \ref{sec:Impact of near horizon flat density profile}, we direct our attention to density profiles that exhibit vanishing characteristics in the vicinity of the BH. Lastly, Section \ref{sec:Discussion and conclusion} wraps up the discussion, drawing conclusions and offering final remarks on our findings and their implications.

In the work, we take the geometrized unit and set $G=c=1$. All the length and mass scales are expressed in terms of the mass of the BH, namely $M_{\rm BH}$. Throughout the paper, we set $M_{\rm BH} =1$, except when it is required to explicitly show $M_{\rm BH}$.

\section{General structure for individual distribution}\label{sec:General structure for individual distribution}

We begin the work by taking established density distribution from the literature. Firstly, we take the following density structure as Ref. \cite{Figueiredo:2023gas}, 

\begin{equation}\label{eq: density not scaled}
\rho(r) = \rho_0\left(\frac{r}{a_0}\right)^{-\gamma}\left[1 + \left(\frac{r}{a_0}\right)^\alpha\right]^{(\gamma - \beta)/\alpha}
\end{equation}

The dependence of the density profile on the coefficients $(\alpha, \beta, \gamma)$ in Eq.  (\ref{eq: density not scaled}) offers a rich spectrum of profiles to explore. These coefficients play distinct roles in shaping the profile: $\beta$ and $\gamma$ govern its behavior at both small and large scales, while $\alpha$ controls the transition's sharpness, making it a versatile tool for modeling various astrophysical systems. Notably, the slope of the profile experiences a notable transition over a characteristic spatial scale defined by $a_0$. In our analysis, we delve into two fundamental models extensively used for understanding the distribution of DM, the Hernquist model and the Navarro-Frenk-White (NFW) model. While we provide a detailed examination of the Hernquist model here, the NFW model is discussed in Appendix \ref{NFW}. The Hernquist profile (Hern) corresponds to specific coefficient values $(\alpha, \beta, \gamma) = (1, 4, 1)$ \cite{Hernquist:1990be}. For the Hernquist profile, the density takes the following form,

\begin{equation}   
    \rho_{Hern} =  \rho_0   \left(\frac{r}{a_0}\right)^{-1}\left[1 + \left(\frac{r}{a_0}\right)\right]^{-3}
\end{equation}

The profile described in Eq.~(\ref{eq: density not scaled}) has a quite generic structure allowing sufficient freedom to model the density by varying the parameters. However, we will not go into such details in the current work.  We will rather consider a different profile, namely, the Einasto model (Ein) \cite{Einasto:1965, Haud:1986yj, Graham:2005xx, Prada:2005mx}. The profile has a different structure from that of in Eq.~(\ref{eq: density not scaled}), given by:

\begin{equation}
    \rho(r)_{Ein} = \rho_e \exp\left[-d_n \left\{\left(\frac{r}{r_e}\right)^{1/n}-1 \right\}\right],
\end{equation}

with $n = 6$, $d_n = 3n-1/3 + .0079/n$ \cite{Graham:2005xx,Prada:2005mx}, and $\rho_e$ representing the density at the radius $r_e$, which defines a volume containing half of the halo mass. To make the paper easy to understand we will rename $r_e$ and $\rho_e$ as $r_e \to a_0$ and $\rho_e \to \rho_0$. For a given density we define $M_{\rm Halo}$ as,

\begin{equation}
 M_{\rm Halo}=   \int_2^{10^7}  4\pi r^2 \rho(r) \,\, dr  ,
\end{equation}

A density profile therefore is completely determined if $a_0$ and $\rho_0$ are known. $M_{\rm Halo}$ is then known in terms of these two parameters. However, we will choose $M_{\rm Halo}$ to be the free parameter of a profile rather than $\rho_0$. Hence, in the current work, we will set the value of $\rho_0$ for all the density profiles by setting the total halo mass up to a certain radius. For this purpose, we use the following,

\begin{equation}
    \rho_0 = \frac{M_{\rm Halo}}{\int_2^{10^7}  4\pi r^2 (\rho(r)/ \rho_0) \,\, dr },
\end{equation}

The properties of a profile are therefore solely determined by setting the values for $(a_0, M_{\rm Halo})$. In the plots, these numbers are demonstrated as $(a_0, M_{\rm Halo})$. The relevant quantity to focus on is $(a_0/M_{\rm BH}, M_{\rm Halo}/M_{\rm BH})$. Since we have set $M_{\rm BH}=1$ our labels will be $(a_0, M_{\rm Halo})$. However, it should always be kept in mind that these values are relative to black hole mass.

\begin{figure*}
\includegraphics[width=85mm]{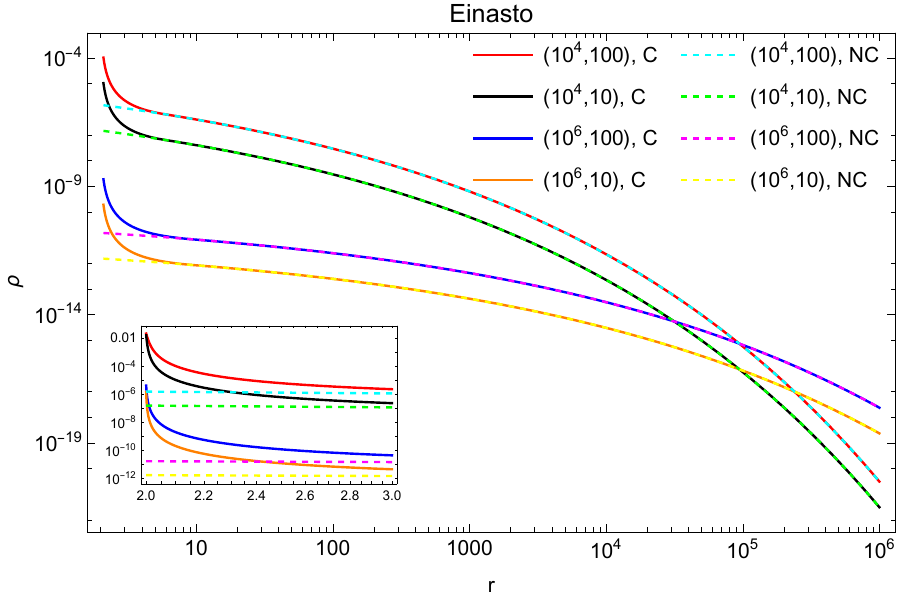}
\includegraphics[width=85mm]{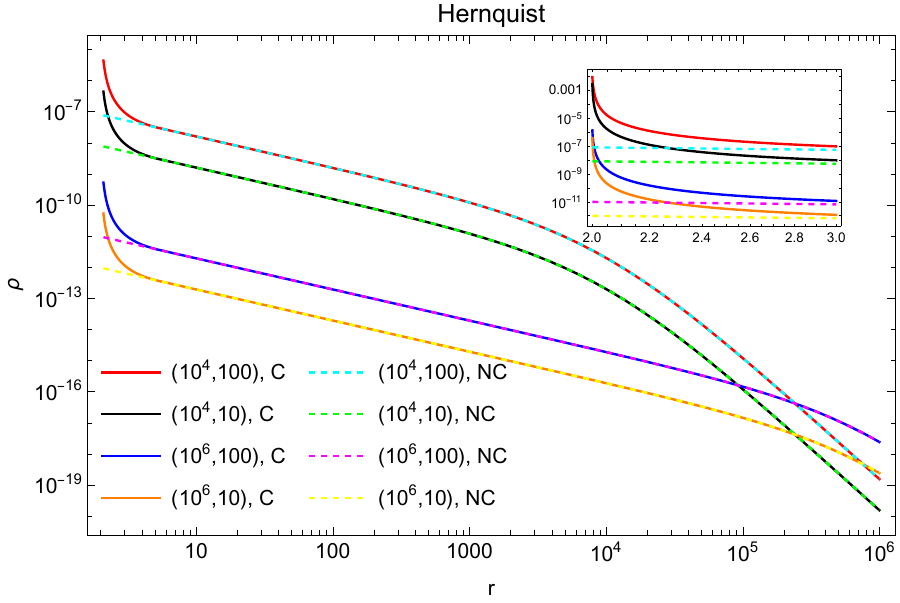}
\includegraphics[width=85mm]{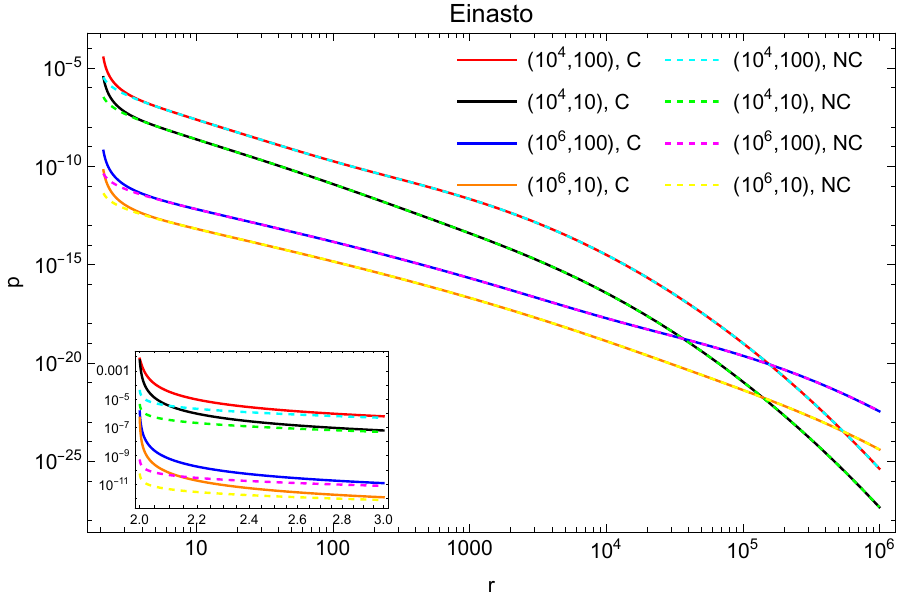}
\includegraphics[width=85mm]{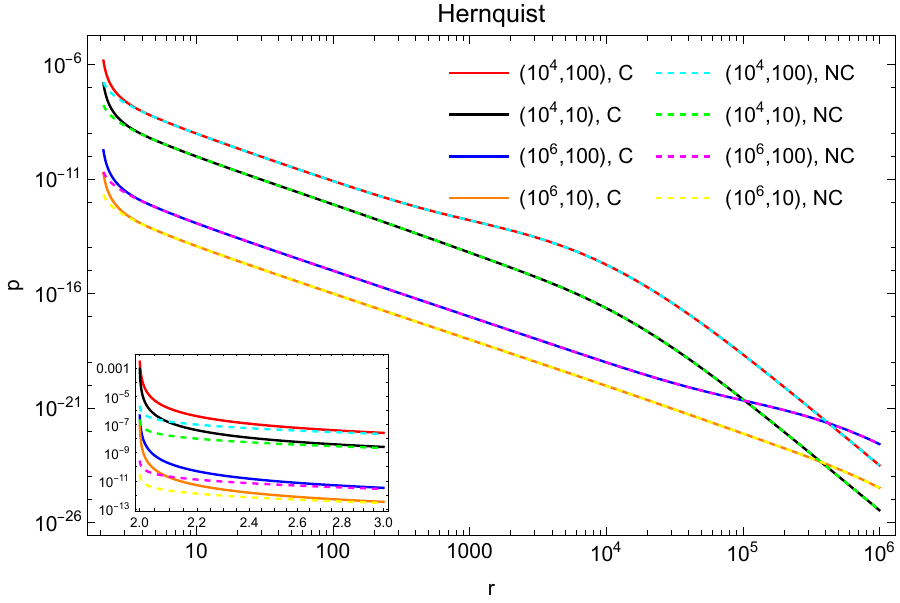}
\includegraphics[width=85mm]{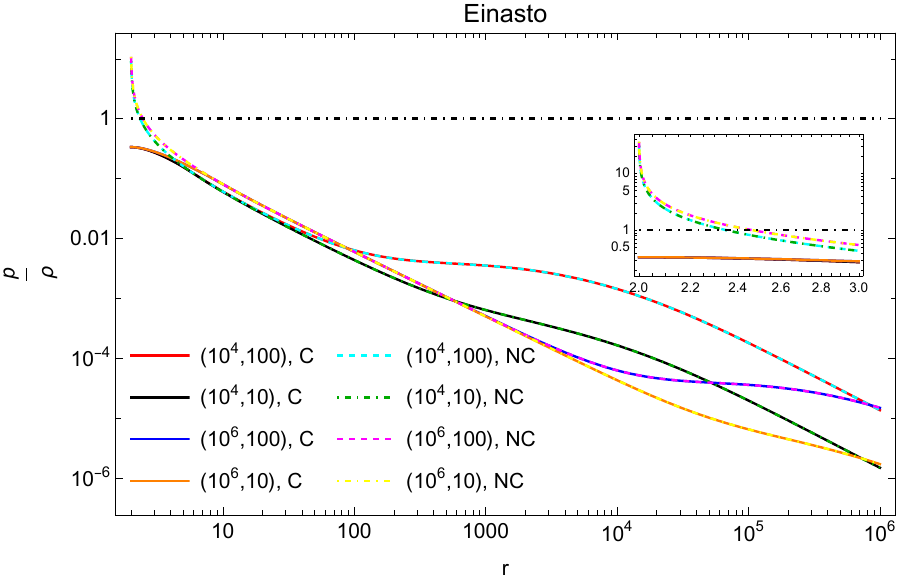}
\includegraphics[width=85mm]{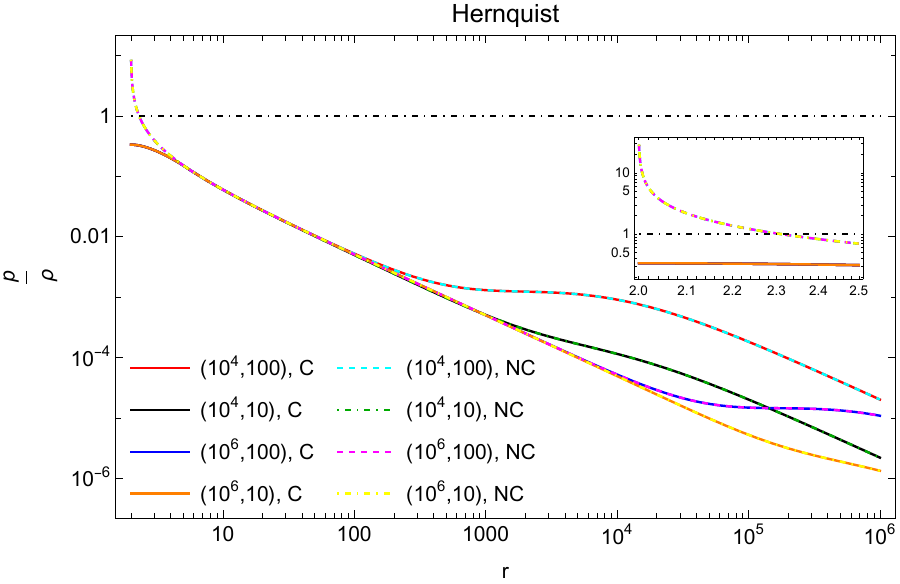}
\caption{In this figure we consider $p_r=p_t=p$. In the first row the density and in the second row the pressure solution of the TOV equation is demonstrated. In the last row, the behavior of the dominant energy condition across the radius is demonstrated. For a given $(a_0, M_{Scale})$ both the C and NC solution is demonstrated. Near the BH the condition gets violated for NC solutions. However, the C solutions do not violate it. All the main plots are demonstrated from $r=2.1$ to $r=10^6$ while in the subplot down to $r=2.001$ is plotted.}
\label{fig:radial pressure density} 
\end{figure*}

\section{Energy condition and Sound speed problem}\label{sec:Sound speed problem}

Our motivation is to investigate in detail the effect of dark matter on black hole geometry. However, the matter distribution itself depends on the nature of the geometry. In this section, we will model the matter contribution through a single-component ideal fluid energy-momentum tensor. The mass density will be identified with the dark matter density. To model the pressure two approaches will be taken. In general, the anisotropic fluid energy-momentum tensor can be expressed as,

\begin{equation}
    T^{\mu}_{\nu} = diag(-\rho,p_r,p_t,p_t)
\end{equation}

We will limit ourselves to different limits of this more general energy-momentum tensor\footnote{It remains to be seen if this approach is appropriate for the collisionless dark matter at least in the phenomenological sense.}. Following the previous works first, we will take $p_t\neq 0$, $p_r=0$. In the second approach, we will consider isotropic pressure similar to the considerations taken for the pressure in compact stars, i.e. $p_t=p_r \neq 0$. We will demonstrate that there is a problem with the dominant energy condition and sound speed in both cases.

\subsection{Effect of matter for vanishing radial pressure}

In this section, we provide a concise summary of the fundamental equations governing a static, spherically symmetric black hole spacetime situated in a medium with a generic density profile $\rho(r)$ with vanishing radial pressure. These results are similar to that in Ref. \cite{Cardoso:2022whc, Figueiredo:2023gas}, which originally applied this framework to investigate binary systems evolving inside a Hernquist-type matter distribution \cite{Cardoso:2021wlq, Figueiredo:2023gas}. Our approach utilizes the Einstein cluster framework to model a stationary black hole, which is surrounded by a collection of gravitating masses \cite{Konoplya:2021ube}. Within this framework, the energy-momentum tensor is represented by,

\begin{equation}
    T^{\mu}_{\nu} = diag(-\rho,0,p_t,p_t)
\end{equation}

Using the Energy momentum conservation equation, the solution for tangential pressure can be found to be,

\begin{equation}\label{eq:pt}
    p_t(r) = \frac{1}{2}\frac{\rho(r) m(r)}{r-2m(r)}
\end{equation}

The knowledge of density profiles provides us with mass profiles. Using them the tangential pressure profile can be computed. From the radial profile of tangential pressure $p_t$ and density $\rho$, we can define a tangential sound speed as follows,

\begin{equation}
\label{eq: pt sound speed}
    c_{st}^2 = \frac{dp_t/dr}{d\rho/dr}.
\end{equation}

In Fig. \ref{fig:tangential cs} we plot $c_{st}^2$ and $p_t/\rho$ for Einasto and Hernquist profile. The dominant energy condition imposes $\rho\geq p_t$. As can be seen from the $p_t/\rho$ plot the dominant energy condition is violated in the near region for all the configurations. It is also noteworthy that in all the cases the sound speed becomes larger than the speed of light around $r\leq 3.5-5$, depending on the profile. This also explains the violation of the energy condition which is connected to the sound speed. This implies that the model starts becoming unphysical near the BH. This problem is not related to the vanishing radial pressure as it persists even in the presence of the non-vanishing radial pressure. In the next section, we will investigate this aspect and how some of the nature of the problem changes if radial pressure is taken into account.

\subsection{Effect of matter for isotropic pressure}

For a matter configuration with isotropic pressure, the energy-momentum tensor takes the following form,

\begin{equation}
    T^{\mu}_{\nu} = diag(-\rho,p,p,p)
\end{equation}

Using the energy-momentum conservation equation the governing equation for the radial pressure can be found, which is similar to the Tolman-Oppenheimer-Volkoff (TOV) equation \cite{Oppenheimer:1939ne},

\begin{equation}\label{eq:matter radial pressure}
    \begin{split}
        -\frac{dp}{dr} =& \frac{(\rho + p)(m(r) + 4\pi r^3 p)}{r(r-2m(r))}\\
    \frac{dm(r)}{dr} =& 4\pi r^2 \rho.
    \end{split}
\end{equation}

Given a density profile, it is straightforward to obtain the mass function by integrating the second equation in Eq. (\ref{eq:matter radial pressure}) 
from $r_{BH}=2$ to radius $r$ with the boundary condition,
\begin{equation}
    \int_{M_{BH}=1}^m dm= \int_{r_{BH}=2}^r 4\pi r^2 \rho dr.
\end{equation}

Given the mass function obtained above, one integrates the first equation in Eq. (\ref{eq:matter radial pressure}) for the pressure. Since the equation is first order, it requires the specification of a single boundary condition to provide a unique solution. At a large radius $r\sim 10^7$ we expect the pressure to be very small. However, due to a lack of knowledge of the equation of state of the matter, particular consistent values are unknown. To get a consistent value we will use the boundary condition that the value of the radial pressure at $r=10^7$ is exactly equal to the value of tangential pressure described by Eq. (\ref{eq:pt}). Therefore we impose,

\begin{equation}
    p(r=10^7) = p_t(r=10^7),
\end{equation}
where for $p_t(r)$ we use Eq. (\ref{eq:pt}). With the boundary condition, it is possible to find the numerical solution of $p$. With this solution, we define sound speed as \footnote{This definition corresponds to adiabatic perturbations and assumes the adiabatic index governing the perturbations is the same as the adiabatic index governing the equilibrium pressure-density relation \cite{1983bhwd.book.....S}.},

\begin{equation}
\label{eq: sound speed}
    c_s^2 = \frac{dp/dr}{d\rho/dr}.
\end{equation}

\begin{figure*}
\includegraphics[width=85mm]{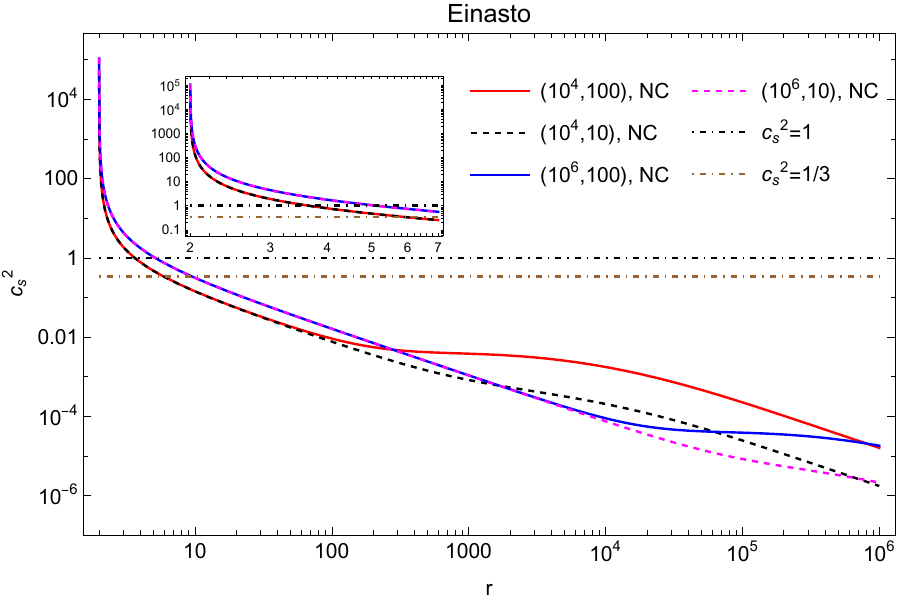}
\includegraphics[width=85mm]{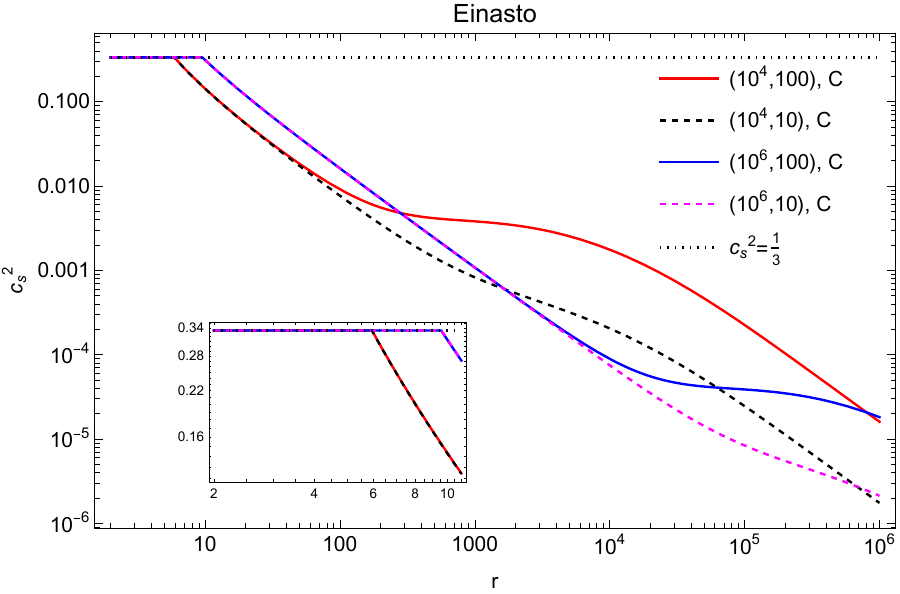}
\includegraphics[width=85mm]{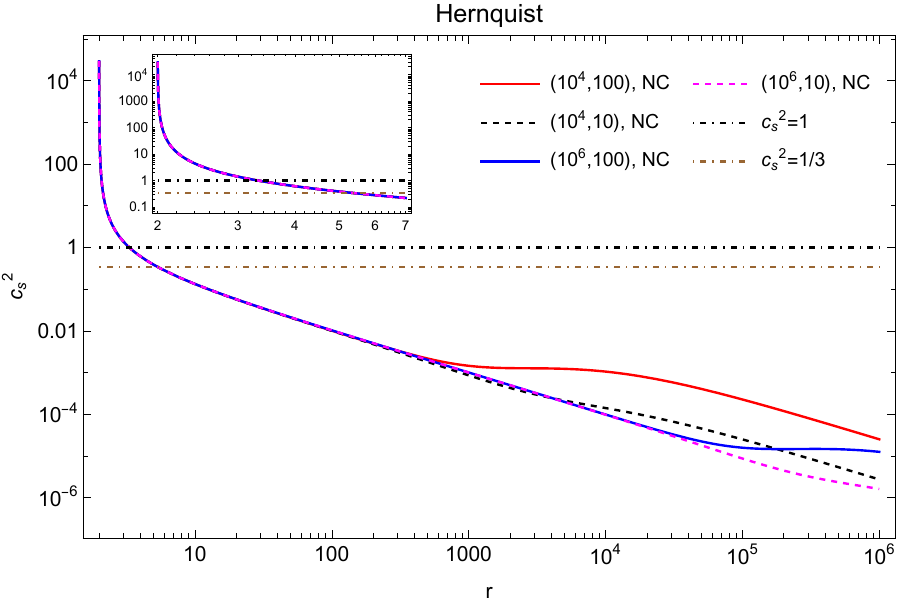}
\includegraphics[width=85mm]{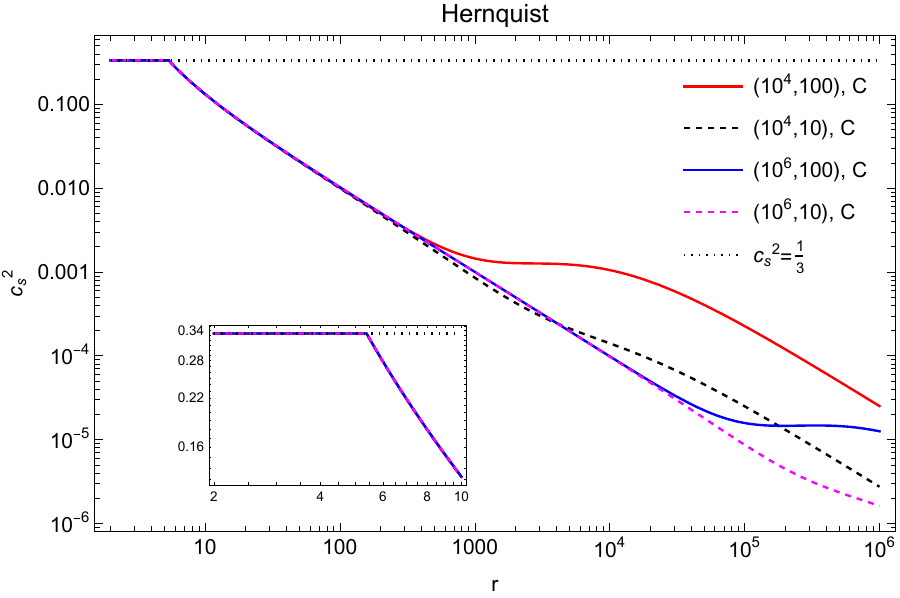}
\caption{In the above figure the sound speed for both the C solution and NC solution is demonstrated. For comparison $c_s^2=1,1/3$ is also plotted. For the NC solution near the BH, the sound speed becomes larger than the speed of light. Due to the condition imposed for the C solutions, the sound speed does not cross the speed of light but there is kink in the radial profile. All the plots are demonstrated from $r=2.001$ to a larger radius.}
\label{fig: cs radial pressure}
\end{figure*}

We will show later that the solution found in this way will result in a similar problem to the case of tangential pressure. The sound speed will become larger than the speed of light. To address this issue we will solve a set of equations simultaneously,

\begin{eqnarray}\label{eq: corrected matter radial pressure}
        -\frac{dp}{dr} =& \frac{(\rho + p)(m(r) + 4\pi r^3 p)}{r(r-2m(r))}
        \\
    \frac{d\rho(r)}{dr} =&  \mathcal{R}(r)\\
    \frac{dm(r)}{dr} =& 4\pi r^2 \rho.
\end{eqnarray}

The function $\mathcal{R}(r)$ is defined as,
\begin{equation}
    \begin{split}
        \mathcal{R}(r) =  \begin{cases}
    \frac{d\rho_{Ein/Hern}(r)}{dr},& \text{if } \left|\frac{d\rho_{Ein/Hern}(r)}{dr}\right|\geq  3\left|\frac{dp(r)}{dr}\right|\\
    3\frac{dp(r)}{dr},              & \text{otherwise}
\end{cases}
    \end{split}
\end{equation}
The reason behind such imposition is to keep the sound speed subluminal as well as satisfy the virial theorem, i.e. $c_s^2\leq 1/3$, which will make the solution satisfy the dominant energy condition. However, this condition will create a kink in the sound speed profile which must be addressed in the future. Note, that this is the key innovation of the current work. This approach is completely different from previous approaches. As will be demonstrated later this innovation makes the sound speed physical.

The boundary condition of pressure is taken to be the same as before, while  the boundary condition of density is taken such that at a large distance density profile matches that of Hernquist or Einasto,

\begin{eqnarray}
        p(r=10^7) =& \,p_t(r=10^7),\\
    \rho(r=10^7) =& \,\rho_{Ein/Hern}(r=10^7).
\end{eqnarray}

\begin{figure*}
\includegraphics[width=85mm]{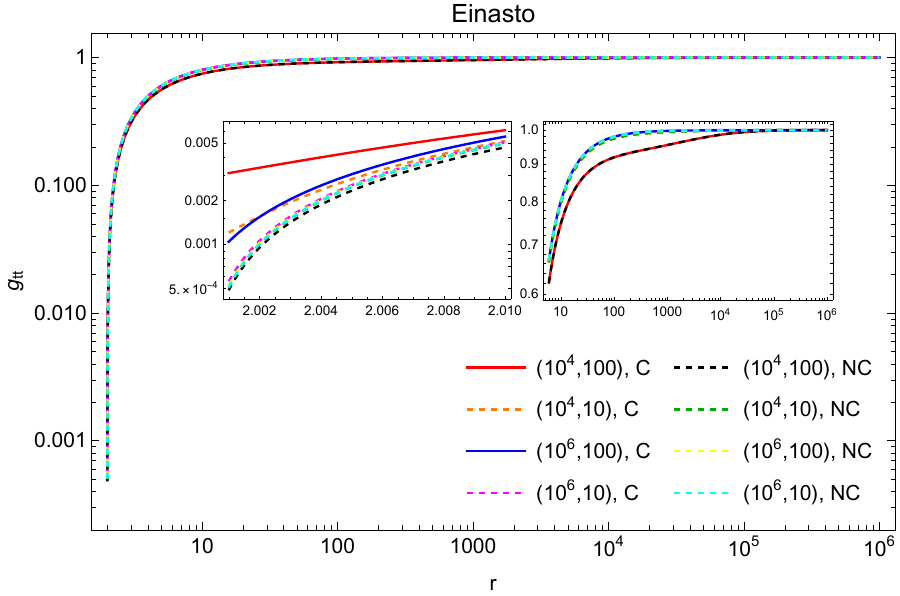}
\includegraphics[width=85mm]{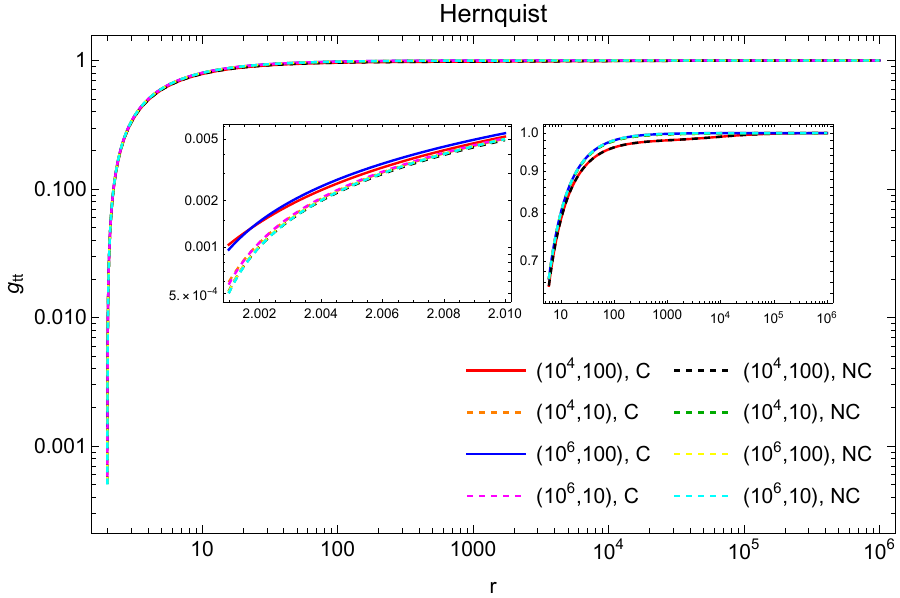}
\includegraphics[width=85mm]{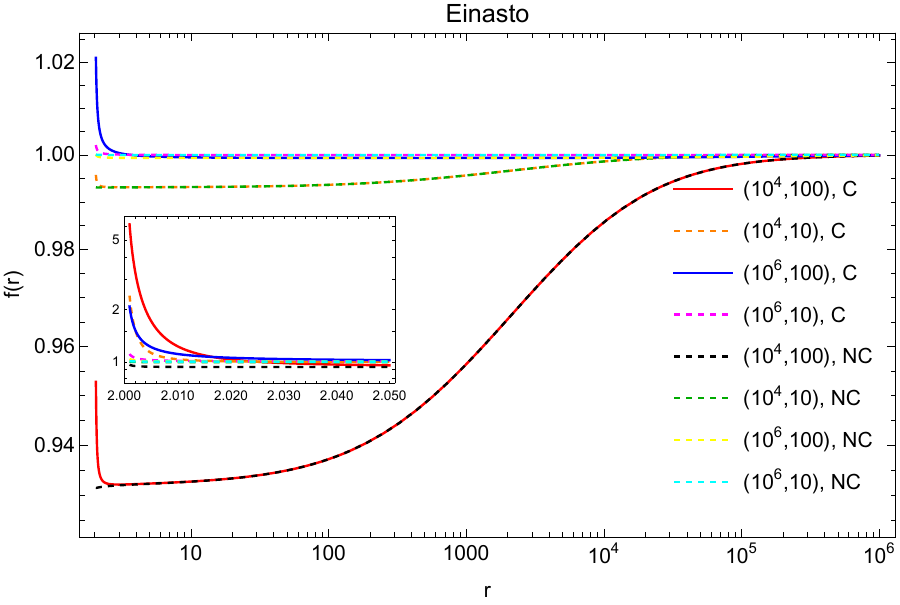}
\includegraphics[width=85mm]{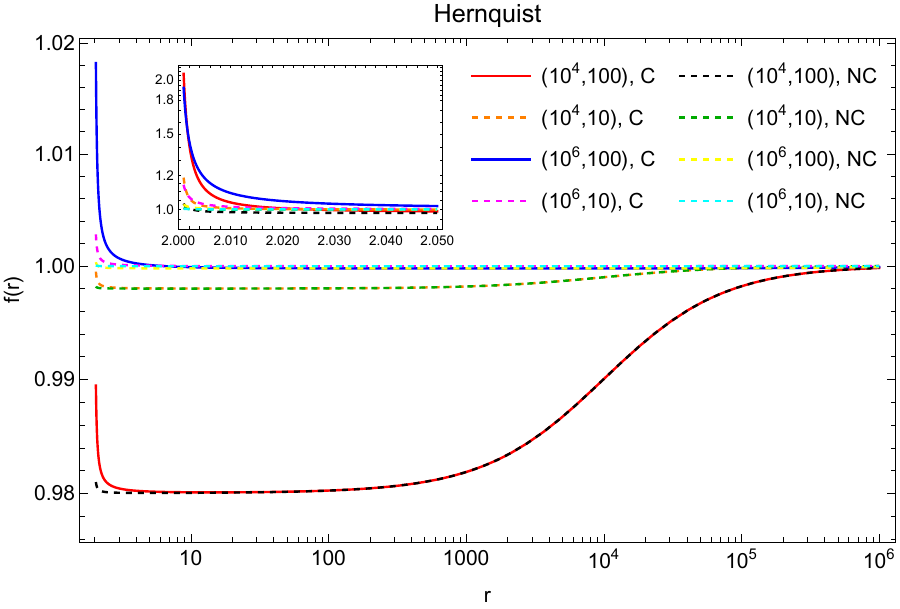}
\includegraphics[width=85mm]{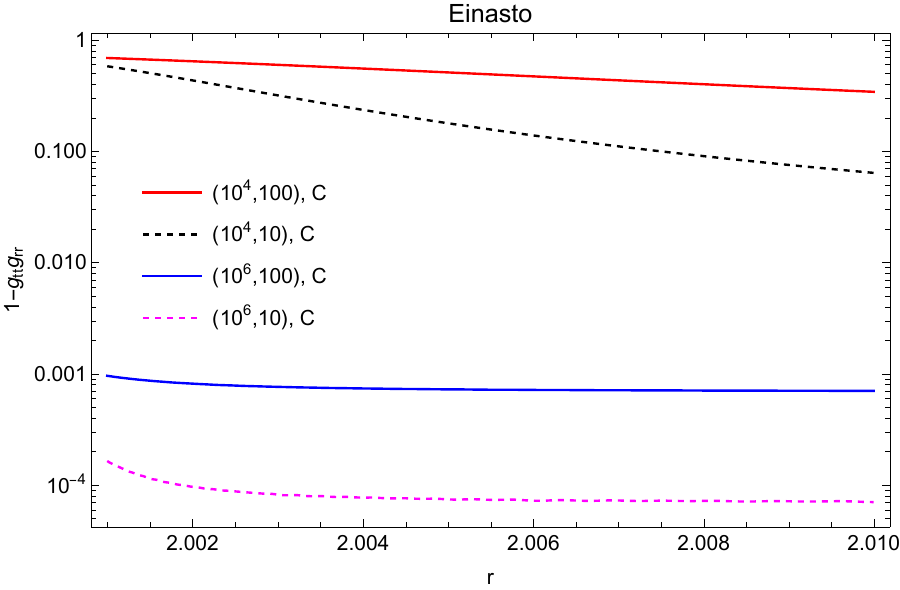}
\includegraphics[width=85mm]{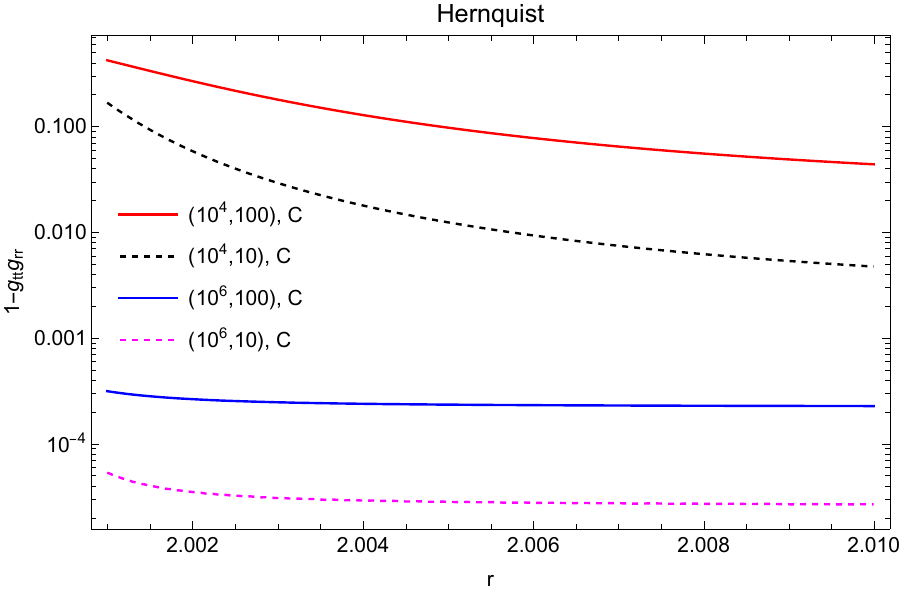}
\caption{In the above figures the solution of the metric is demonstrated. In the first row metric component $g_{tt}$ is shown. All the metric plots are demonstrated from $r=2.001$ to $r=10^6$ to demonstrate the structure for individual profiles. However, in the right subplots from $r=6$ to $r=10^6$ is shown and in the left subplots from $r=2.001$ to $r=2.1$ is shown. In the second row, the deviation function $f(r)$ is shown. All the deviation function plots are demonstrated from $r=2.05$ to $r=10^6$ while in the subplot, from $r=2.001$ to $r=2.05$ is shown. In all the cases the metric goes to zero rapidly near $r\sim 2$. The deviation function increases near $r\sim 2$. The increase is more for the C solution compared to the NC solution. Despite this increase, the $g_{tt}$ component takes a small value near $r\sim 2$. In the final row $1-g_{tt}g_{rr}$ has been plotted. For a Schwarzschild BH even near the horizon, $g_{tt}g_{rr}$ stays equal to unity. In the current case, although this factor deviates from the BH value of 1, the deviation is not extremely large.}
\label{fig:metric}
\end{figure*}

\begin{figure*}
\includegraphics[width=85mm]{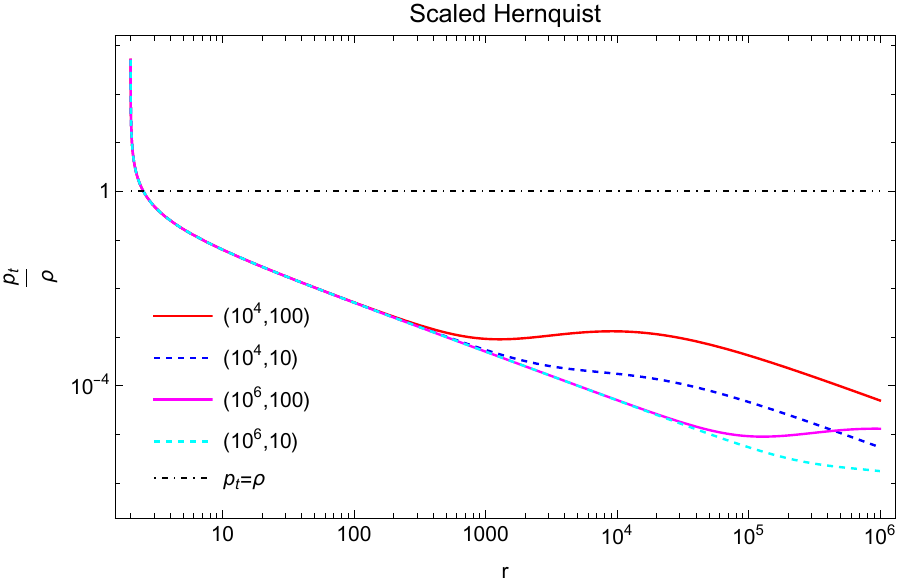}
\includegraphics[width=85mm]{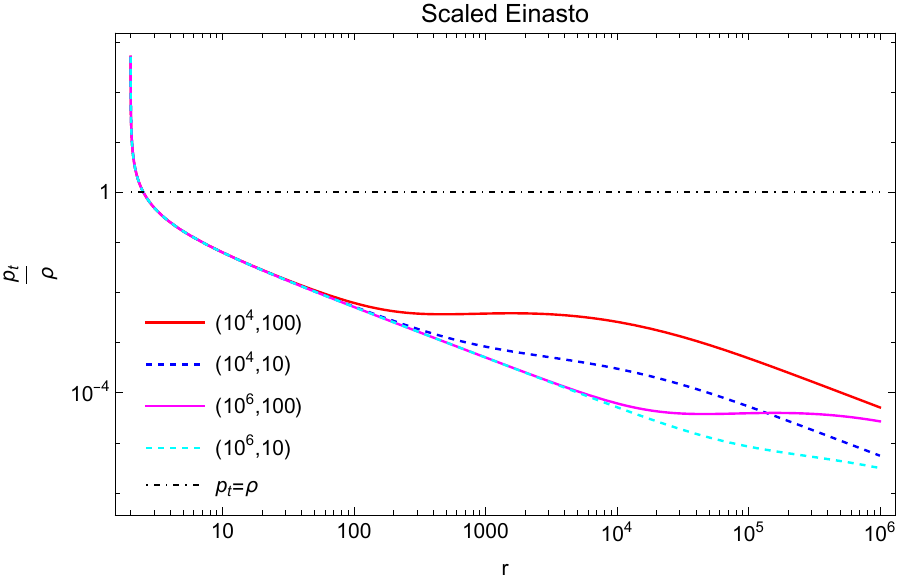}
\includegraphics[width=85mm]{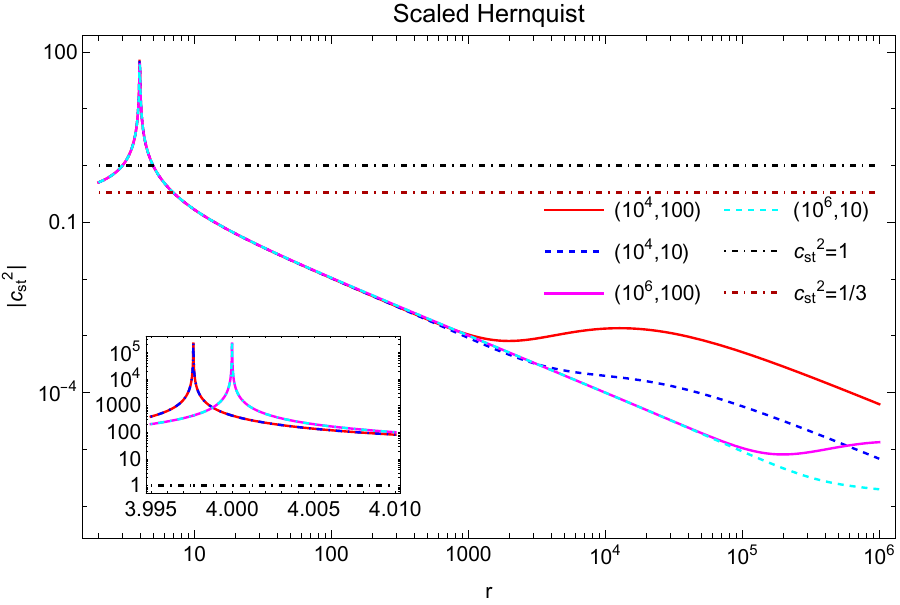}
\includegraphics[width=85mm]{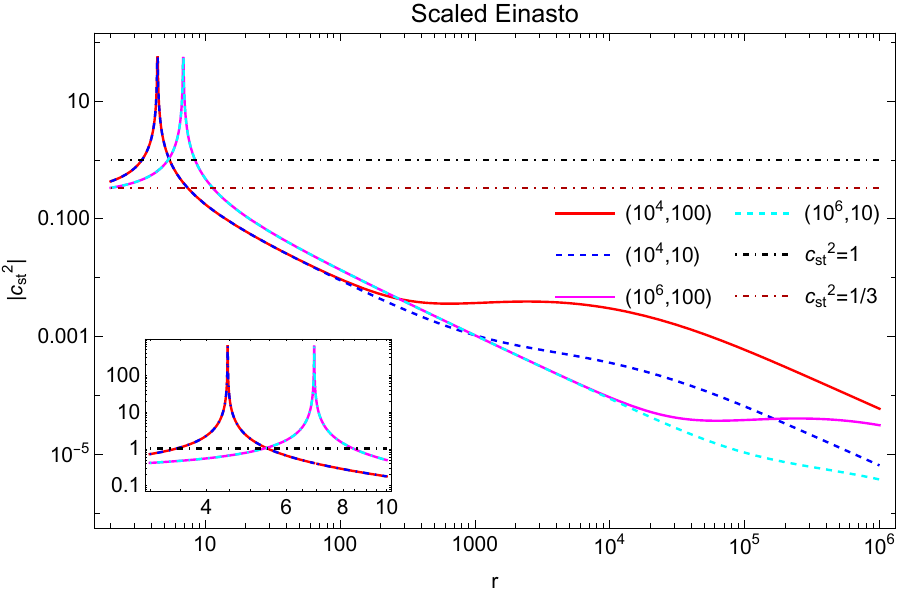}
\caption{Here we show the results of scaling different density profiles by $(1-2/r)$. In the first row of the above plots, we demonstrate how the dominant energy condition varies across the radius for the density profiles which vanishes near the BH. The condition is clearly violated in the near zone. In the last row, we show the behavior of the magnitude of the tangential sound speed. We show the magnitude as $c_{st}^2$ becomes negative below the divergence point. All the profiles show diverging behavior in the near BH region. This feature is unique to this kind of density profile where the density has a maxima.}
\label{fig: scaled DEC and cst}
\end{figure*}

\begin{figure*}
\includegraphics[width=85mm]{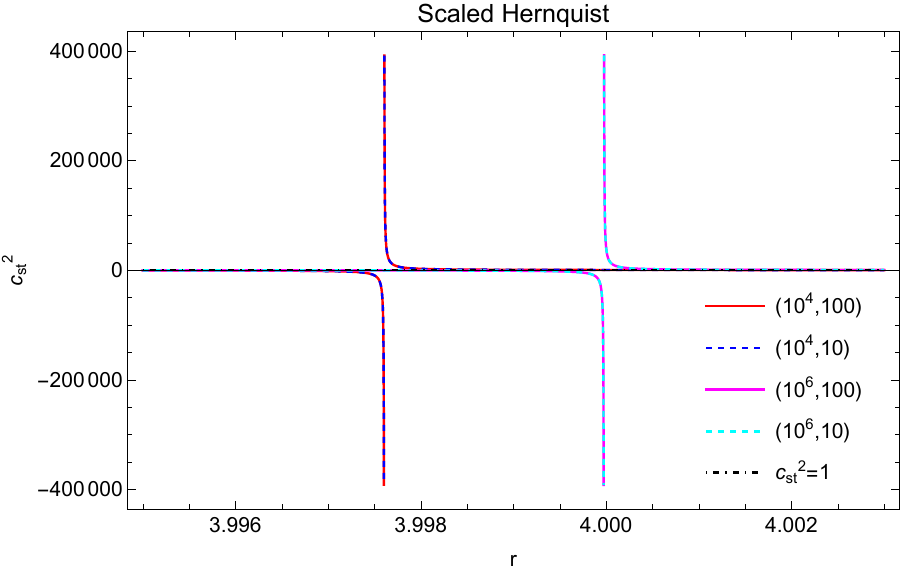}
\includegraphics[width=85mm]{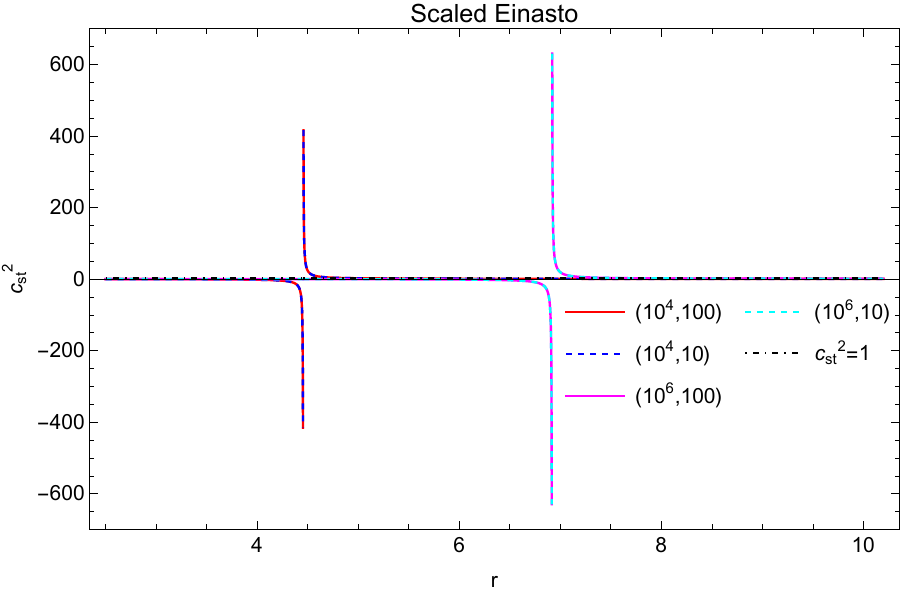}
\includegraphics[width=85mm]{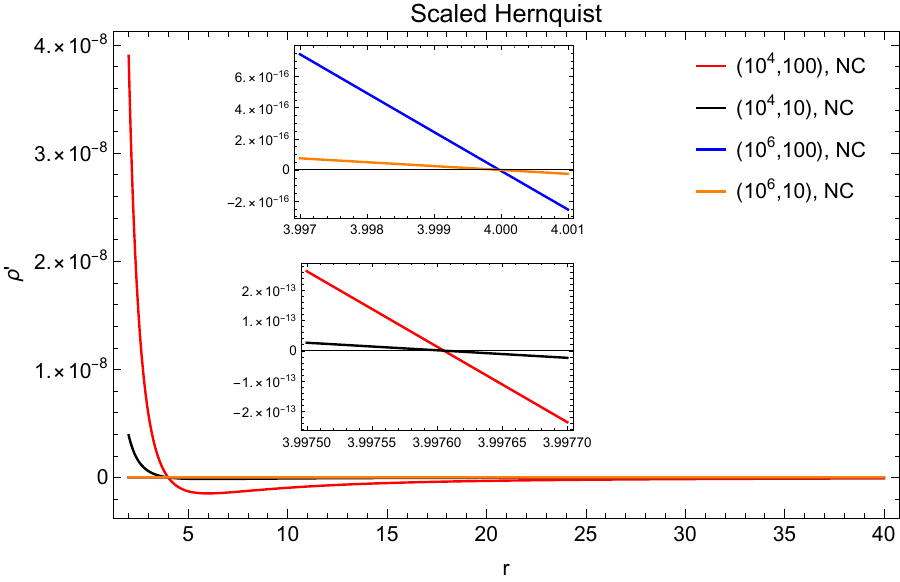}
\includegraphics[width=85mm]{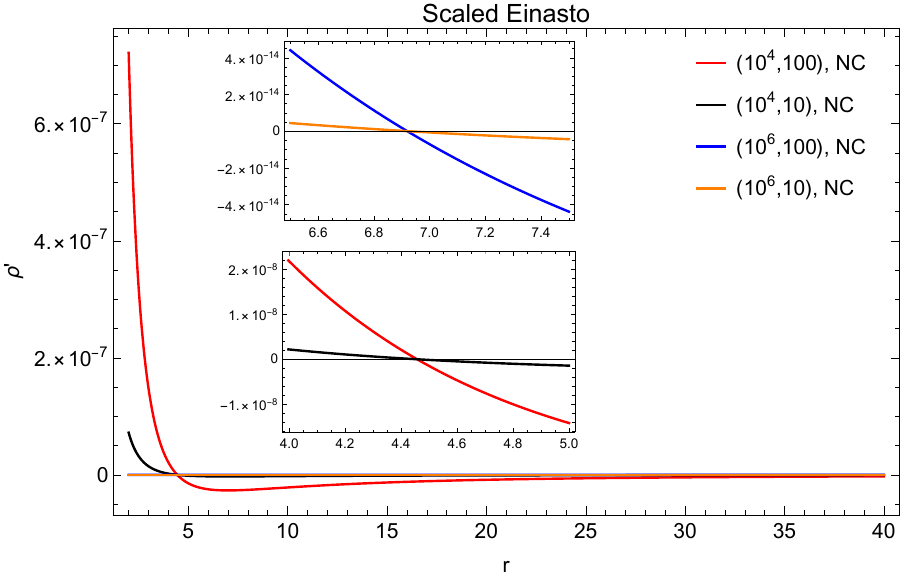}
\caption{In the first row we plot $c_{st}^2$ and we zoom in the diverging region. For all the profiles, below a certain radius $c_{st}^2$ becomes negative. In the second row, we plot the derivative of the density function with respect to the radius. We find that in most of the range of radius, the derivative is negative and in the near zone it becomes positive. The point when it becomes zero corresponds to the maxima of the density function. As a consequence $c_{st}^2$ blows up at that point and becomes negative for a smaller value of radius. This indicates that the maxima of the density is directly connected to the unphysical behavior of the sound speed.}
\label{fig:scaled rho prime}
\end{figure*}

The solutions found in this way are labeled as {\it corrected solution} (C) while the solutions from the previous approach are labeled as {\it not corrected solution} (NC), where {\it not corrected} refers to the not corrected energy condition and sound speed. In Fig. \ref{fig:radial pressure density} we plot both the C and NC solutions. We also plot $p/\rho$. As it can be noticed, $p/\rho$ violates the dominant energy condition for NC solutions, whereas for C solutions the energy condition is satisfied. Interestingly the density and pressure for C solutions increase more with decreasing radius compared to the NC solutions. Therefore a matter profile that is consistent with causality seems to prefer an increased density near a BH rather than a decrease in density after reaching maxima near the BH. This aspect will be discussed more in later sections.

In Fig. \ref{fig: cs radial pressure} we show the behavior of the sound speed for both the C and NC solutions. The sound speed for NC solutions for all the profiles becomes larger than the speed of light in the near region. The C solution though does not violate physicality, it demonstrates a kink. This kink arises solely from the hard-cut off. A better cut-off needs to be found to address this issue. However, the key point is that by imposing a physical condition on the sound speed the resulting density structure in the near horizon zone becomes quite larger which is contrary to the expectation that the density should vanish in the near zone. This aspect will be discussed in more detail in later sections.

\section{Metric solutions with DM profile}\label{sec:Metric solutions with DM profile}

In the last sections, we studied the matter profiles in detail for both the vanishing and nonvanishing radial pressure. We also explored the properties of sound speed and the impact of making it physical. With the Density and pressure profile at hand, we compute the metric components in this section. We will focus only on the case where the radial pressure does not vanish. The results for vanishing radial pressure can be found in Ref. \cite{Figueiredo:2023gas}.

\subsection{Metric solutions}

\begin{figure*}
\includegraphics[width=85mm]{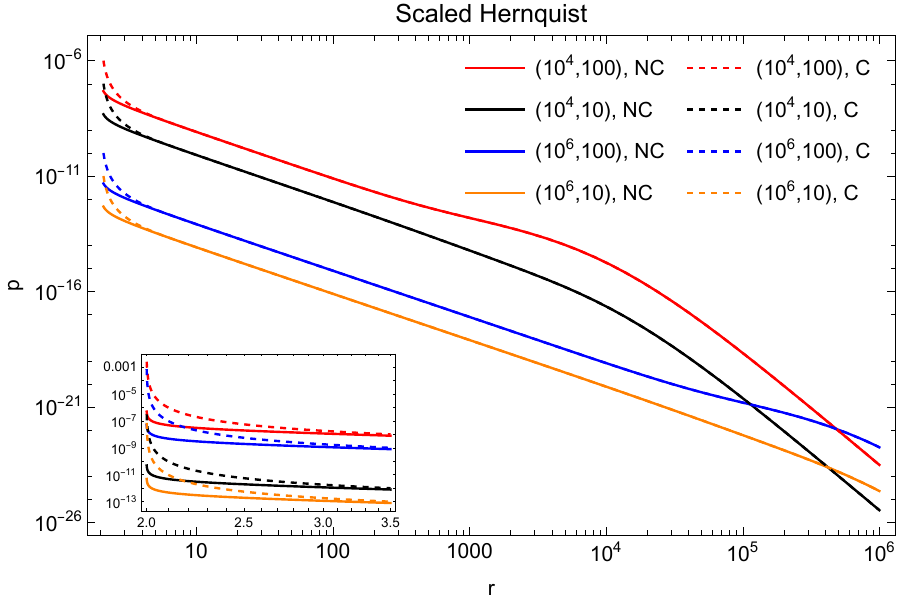}
\includegraphics[width=85mm]{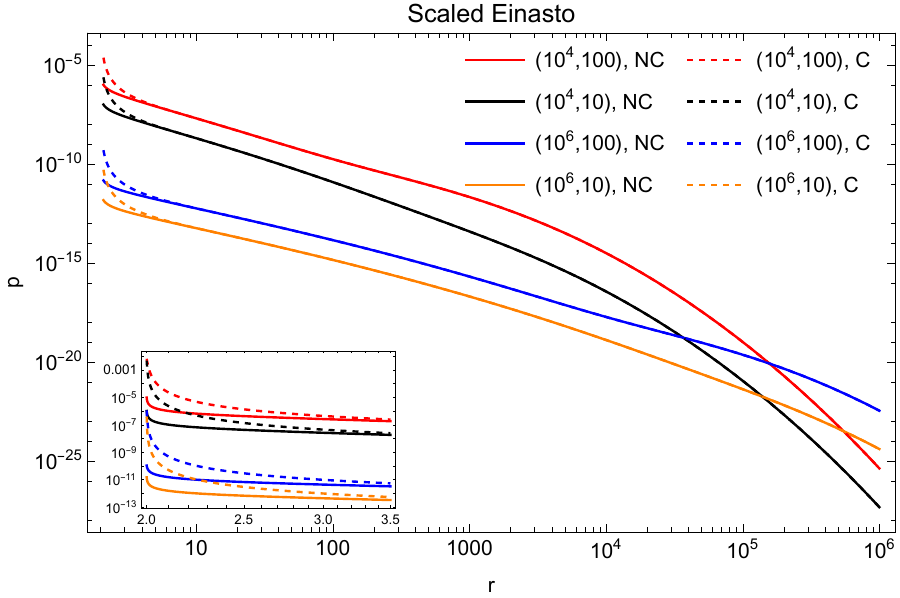}
\includegraphics[width=85mm]{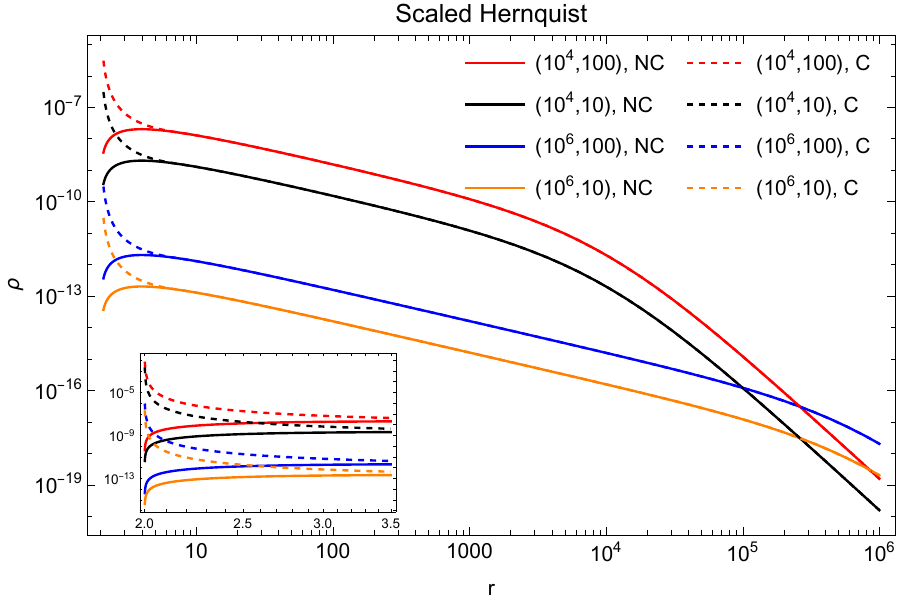}
\includegraphics[width=85mm]{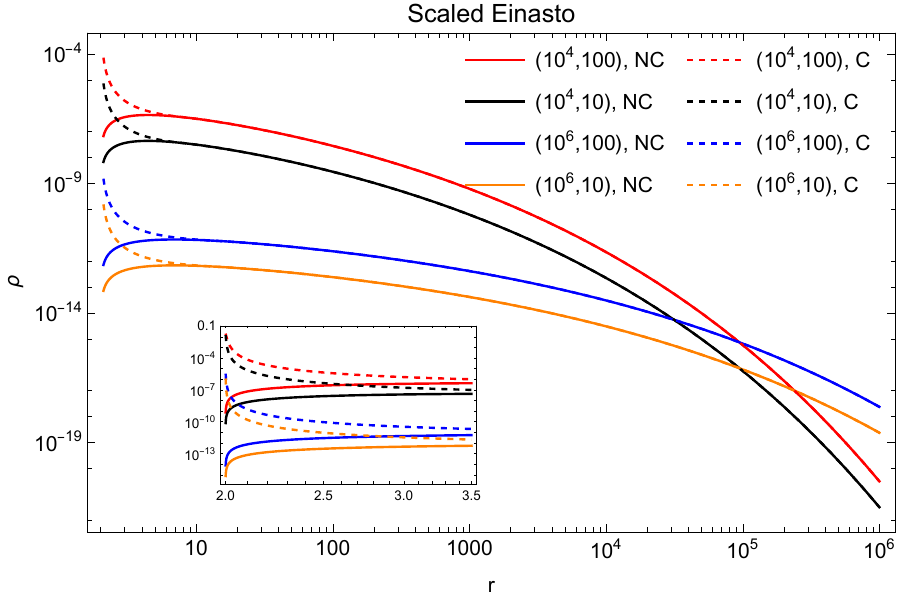}
\caption{In the first row of the above we demonstrate the pressure profile found by solving the TOV equation with the scaled density solution. Pressure in the C solution shows a steeper growth compared to the NC solution. In the second row, we demonstrate the density profile for the same. NC solutions show a maxima and decaying structure of density. The C solutions demonstrate an increase in density profile which is significantly different from the NC solution.}
\label{fig:scaled density}
\end{figure*}

A static spherically symmetric metric can be described in the following manner,

\begin{equation}
    ds^2 = -e^{\nu(r)}dt^2 + e^{\lambda(r)} dr^2 + r^2 (d\theta^2 + \sin^2\theta d\phi^2) 
\end{equation}

Using the metric and the energy-momentum tensor in the Einstein equation it can be shown,

\begin{equation}
\label{eq: nu eq}
    \frac{d\nu}{dr} = -\frac{2}{\rho + p} \frac{dp}{dr}.
\end{equation}

\begin{equation}
    \begin{split}
        e^{\lambda} = \frac{r}{r-2m(r)}.
    \end{split}
\end{equation}

To find the solution for $\nu(r)$ for a given matter profile we require a boundary condition. We impose that $e^{\nu(10^7)}=1$. With the boundary condition, it is straightforward to find the solution for the metric. In the case of Schwarzschild BH in vacuum $e^{\nu(r)}=1-2/r$. In the presence of matter, the metric solution should deviate from the vacuum solution. To quantify this deviation we define,

\begin{equation}
\label{eq: f func}
    f(r) = \frac{e^{\nu(r)}}{(1-2/r)}.
\end{equation}

In Fig. \ref{fig:metric} we demonstrate the metric solutions $g_{tt} = e^{\nu(r)}$. Notably in all the cases near $r\sim 2$ the metric rapidly falls off to small values, which implies the existence of a horizon where the metric component $g_{tt}\rightarrow 0$. We also plot the deviation function $f(r)$ in the second row. Note, that the deviation function seems to increase near the horizon. The increase is more prominent for the C solutions compared to the NC solutions. However, despite the increase in the deviation function the metric component $g_{tt}$ as a whole decreases with decreasing radius.

\subsection{Coordinate singularity makes radial derivatives diverge}

For a vacuum BH, $r=2$ is known to host a coordinate singularity where $g_{tt}$ goes to zero and $g_{rr}$ diverges. From the metric solutions demonstrated earlier similar behavior is demonstrated in the presence of the matter also. It may therefore imply that $dp/dr$ should diverge due to the denominator in Eq. (\ref{eq:matter radial pressure}) at $r=2m(r)$. This as a result can have an impact on the sound speed calculations. However, we would argue that the singularity in Eq. (\ref{eq:matter radial pressure}) is just a coordinate singularity. Hence, similar to the vacuum BH space-time it is also not physical singularity. One should note that as we are outside a BH we should expect $r$ co-ordinate to be pathological at $r\to 2m(r)$. However, if we take the equivalent of the Tortoise coordinate, $\frac{dr^*}{dr} =e^{-\nu(r)}$, the equation simplifies to,

\begin{equation}
    -\frac{dp}{dr^*} = \frac{(\rho + p)(m(r) + 4\pi r^2 p)}{r^2}e^{\nu(r)}e^{\lambda(r)}
\end{equation}

Note that the divergence in the derivative of pressure arises solely from $e^{\lambda(r)}$ term above. In vacuum BH case $e^{(\nu(r)+\lambda(r))}=1$. In the third row of Fig. \ref{fig:metric} we show that in the current case also $e^{(\nu(r)+\lambda(r))}\sim 1$. Which again implies the singularity is just a coordinate singularity. Hence, the pressure derivative with respect to the tortoise radius does not diverge. Notably, the values of sound speed therefore stay unaffected even if defined with respect to $r^*$ as,

\begin{equation}
    c_s^2 = \frac{dp/dr^*}{d\rho/dr^*}=\frac{dp/dr}{d\rho/dr}.
\end{equation}

Therefore the unphysical nature of the sound speed described in the earlier sections is not due to the coordinate singularity. The unphysical nature solely arises from the description of the matter distribution.

\section{Impact of near horizon maxima of density profiles}\label{sec:Impact of near horizon flat density profile}

In the previous sections, we considered Einasto and Hernquist profiles and studied the matter distribution, energy conditions, sound speed, and metric structure. We also demonstrated that the physical sound speed condition introduces an increased overdensity region near the black hole.

With both the Newtonian and relativistic analyses it has been found that the density profiles with a BH at their core exhibit a vanishing density at the horizon and develop a cusp with a lengthscale determined by the BH's mass \cite{Sadeghian:2013laa, Gondolo:1999ef}. The specific mathematical form of the profile determines the detailed nature of the profiles (i.e. the slope, maximum value, etc.), which could have implications for accurately modeling the GW signals emitted by coalescing binaries \cite{Speeney:2022ryg}. To assess the impact of the vanishing density near BH we introduce rescaling the density profile according to $\rho(r) \rightarrow \rho(r)(1 - 2 /r)$, following the results of \cite{Cardoso:2020iji, Figueiredo:2023gas} (check Appendix \ref{app: scale 4} for a different scaling).

In this section, we will demonstrate that such profiles have even more severe sound speed problems compared to the ``normal" Hernquist and Einasto profiles. The density for Einasto is,

\begin{equation}
    \rho_{Ein,S} = \left(1-\frac{2}{r} \right) \rho_e \exp\left[-d_n \left\{\left(\frac{r}{r_e}\right)^{1/n}-1 \right\}\right]
\end{equation}

For the Hernquist profile we take the mass function described in Ref. \cite{Cardoso:2022whc},

\begin{equation}
    m_{Hern,S} = M_{\text{BH}} + \frac{Mr^2}{(a_0 + r)^2} \left(1 - \frac{2M_{\text{BH}}}{r}\right)^2,
\end{equation}
where the label $S$ creates a distinction with the original corresponding distribution from these distributions where densities are scaled to make it vanish near BHs.

\begin{figure*}
\includegraphics[width=85mm]{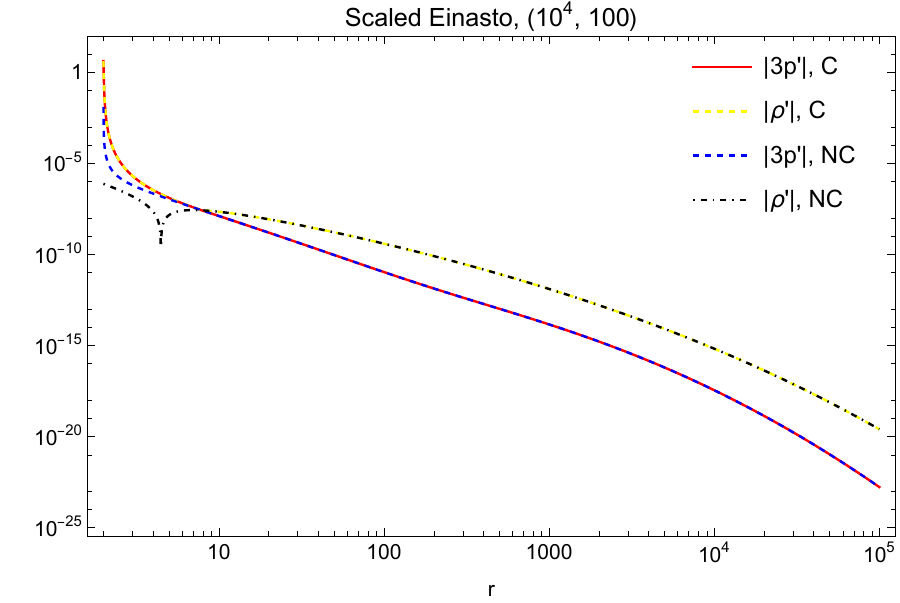}
\includegraphics[width=85mm]{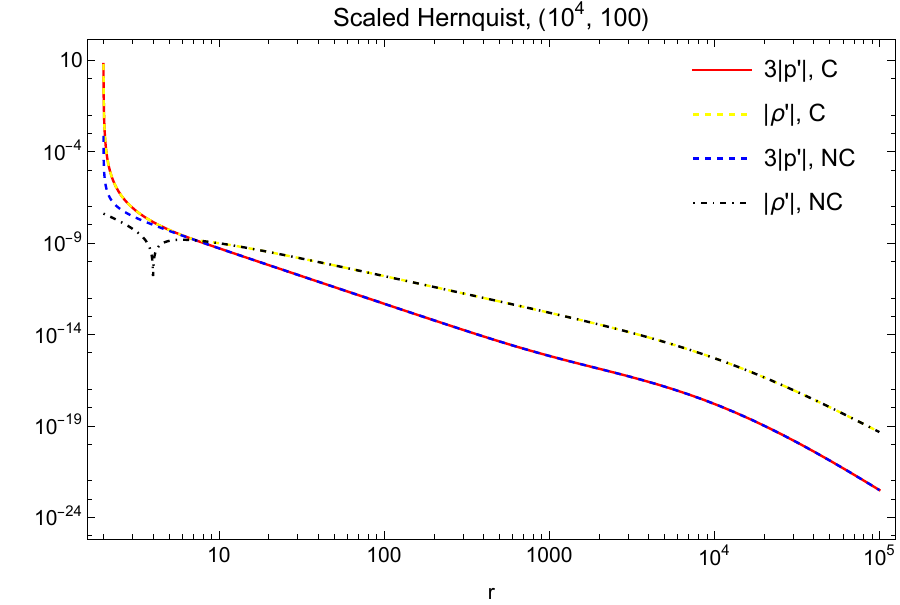}
\includegraphics[width=85mm]{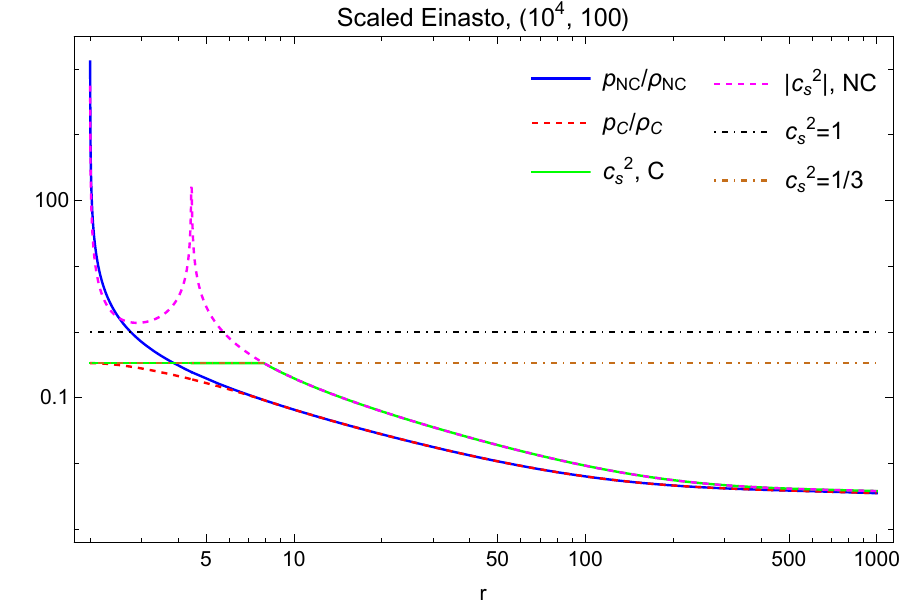}
\includegraphics[width=85mm]{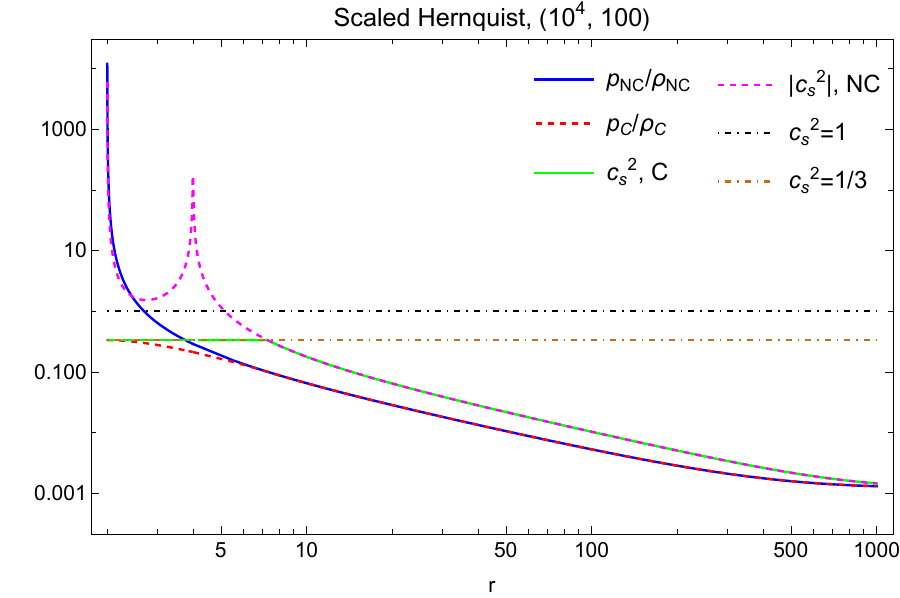}
\caption{In the first row of the above figure we show the behavior of the magnitude of the pressure and density derivative for a $(10^4,\,100)$ configuration for both the scaled Einasto and scaled Hernquist profile. We show both the C and NC solutions. From the structure, the behavior of sound speed is understandable. In the bottom row, we show the dominant energy condition and sound speed behavior. For NC solutions we are showing the magnitude of the sound speed as it becomes negative in the near BH region. Although for the NC solution, both the dominant energy condition and the sound speed are unphysical, for the C solution they both are physical. The kink (in green) in the sound speed requires further investigation. This green curve merges with the NC sound speed (magenta dashed curve) for a larger radius and merges with the brown dot-dashed line representing $c_s^2 =1/3$ for the smaller radial values.}
\label{fig:corrected cs for scaled}
\end{figure*}

\subsection{Matter distribution}

\begin{figure*}
\includegraphics[width=85mm]{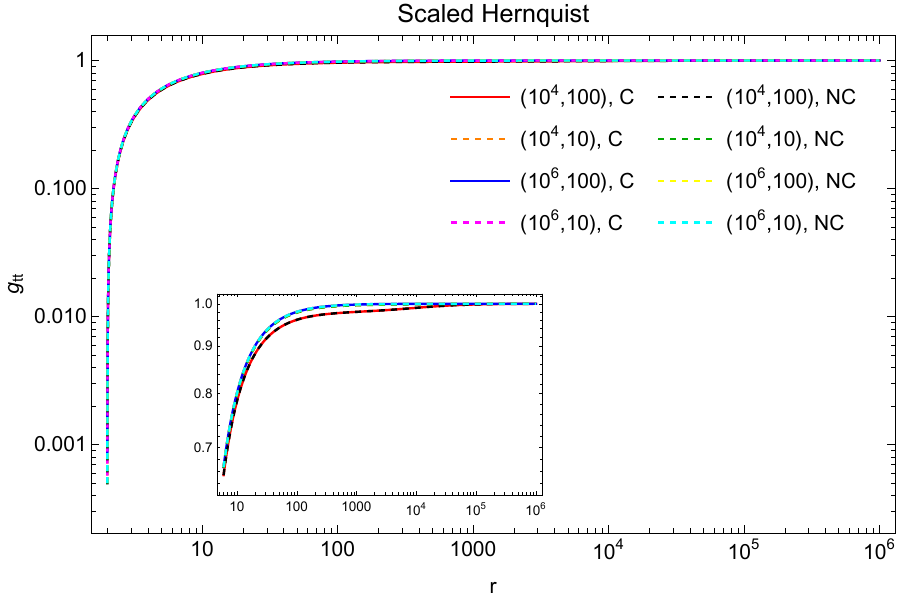}
\includegraphics[width=85mm]{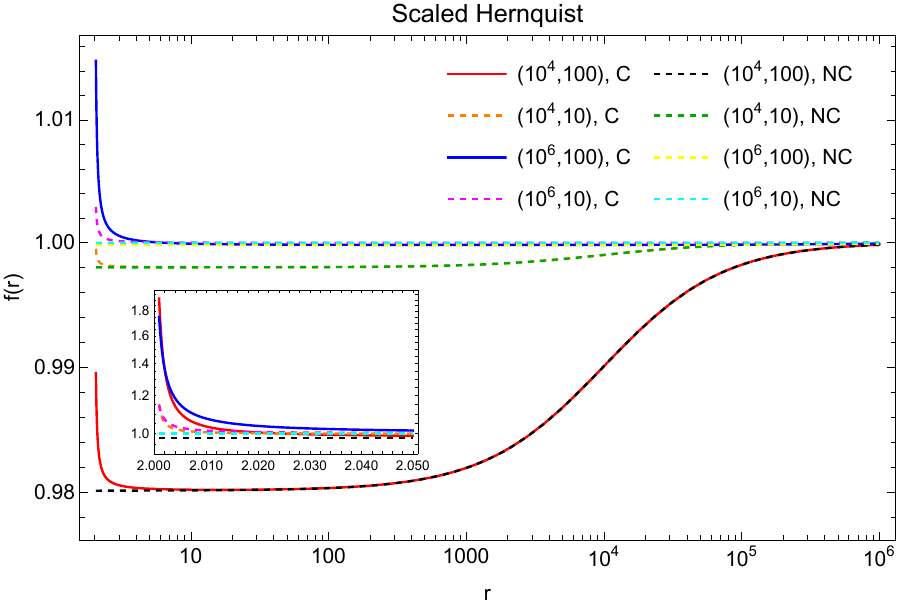}
\includegraphics[width=85mm]{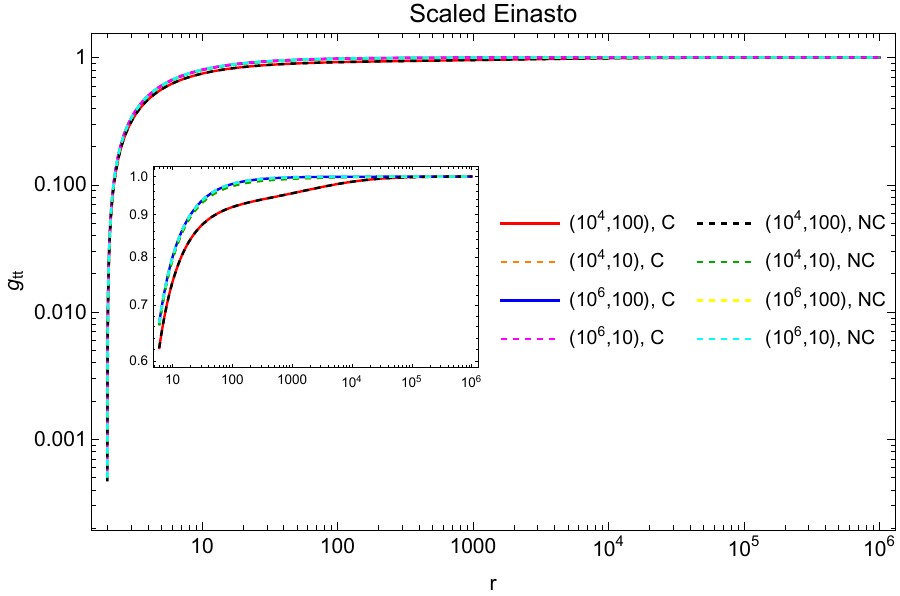}
\includegraphics[width=85mm]{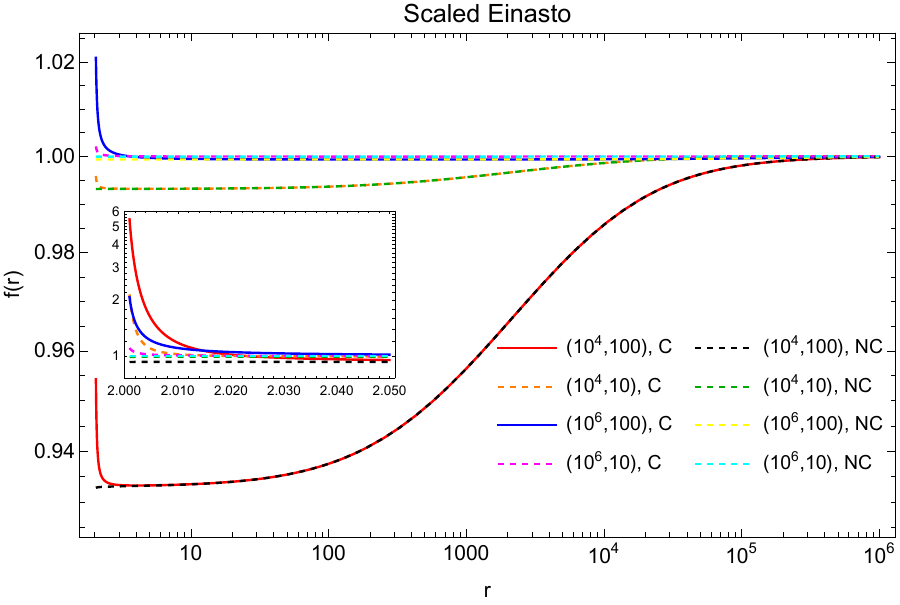}
\caption{In the above figures the solution of the metric is demonstrated for density profiles that vanish near the BH. In the left column metric component $g_{tt}$ is shown. All the metric plots are demonstrated from $r=2.001$ to $r=10^6$ to demonstrate the structure for individual profiles. However, in the subplot, from $r=6$ to $r=10^6$ is shown. In the right column, the deviation function $f(r)$ is shown. All the deviation function plots are demonstrated from $r=2.05$ to $r=10^6$, while in the subplot from $r=2.001$ to $r=2.05$ is shown. In all the cases $g_{tt}$ goes to zero rapidly near $r\sim 2$. The deviation function increases near $r\sim 2$. The increase is more for the C solution compared to the NC solution. Despite this increase, the $g_{tt}$ component rapidly goes to vanishingly small value near $r\sim 2$.}
\label{fig: metric in scaled}
\end{figure*}

With the density and mass function at hand, it is straightforward to compute tangential pressure and corresponding tangential sound speed (Eq. (\ref{eq:pt}) and (\ref{eq: pt sound speed})) when radial pressure is assumed to be vanishing. In Fig. \ref{fig: scaled DEC and cst} in the first row we demonstrate $p_t/\rho$ which violates the dominant energy condition like before. In the second row, we demonstrate the behavior of $|c_{st}^2|$.  We find a divergent nature that is unique to these density structures. In the first row of Fig. \ref{fig:scaled rho prime} we show $c_{st}^2$ near the divergent region. With decreasing radius sound speed starts to grow and diverge. For even smaller values of radius, the square of the sound speed becomes negative.

The origin can be understood by noticing the final row in Fig. \ref{fig:scaled rho prime}. In the final row, we plot $d\rho/dr$. Although in most of the region it stays negative, eventually it starts to go towards the positive direction. In this region, as the derivative is negative the $c_{st}^2$ stays positive. Eventually, $d\rho/dr$ vanishes when the density reaches the maxima and afterward, the derivative becomes positive rendering $c_{st}^2$ negative. At the point where the maxima are reached (say $r_m$) $c_{st}^2$ blows up as the $d\rho/dr$ is in the denominator. However as ${r\to r_{m^+}}$ the derivative is a negative number and as ${r\to r_{m^-}}$ it is a positive number (where $r_{m^-}=r_m -\epsilon$ and $r_{m^+}=r_m+\epsilon$ with $\epsilon\to 0$) making $c_{st}^2$ take respectively the positive and negative values. Therefore the divergence arises solely from the maxima of the density function. The negative values of $c_{st}^2$ arise because after $r=r_m$ the density starts to decrease. Divergent behavior is present for all the profiles considered. Similarly, for all the profiles the density derivative crosses through the zero value. It points towards the possibility that under the single component ideal fluid description of DM, outside a BH density can not reach a maxima, otherwise, sound speed will become unphysical. It may also imply that the assumed energy-momentum structure is inappropriate for the DM. This needs further investigation.

In Fig. \ref{fig:scaled density} we demonstrate the C solution of the TOV equations along with the NC solutions of density and pressure. It is very prominent here that the C solutions do not allow a decreasing density. On the contrary, it makes the density grow near the BH. It is understandable as the maxima of the density profile is directly connected with the diverging sound speed. Therefore the imposition of $0<c_s^2<1/3$ does not allow either the maxima or the decrease in density after the maxima.

In the first row of Fig. \ref{fig:corrected cs for scaled} we show the behavior of the magnitude of the pressure and density derivatives for a $(10^4,\,100)$ configuration for both the scaled Einasto and scaled Hernquist profile. We show both the C and NC solutions. From the structure, the behavior of sound speed is understandable. In the bottom row, we show the dominant energy condition and sound speed behavior. The other $(a_0, M_{\rm Scale})$ configurations also show similar behavior. For brevity, we are not showing them here. Note, C solutions satisfy the energy condition and the sound speed stays physical. Whereas the NC solutions violate the energy condition and the sound speed becomes unphysical. For NC solutions we have plotted $|c_s^2|$ as it becomes negative for lower values of radius.

\subsection{Structure of metric solutions}

With the computed density and pressure we calculate the metric components. The governing equation used is Eq. (\ref{eq: nu eq}). Like before we set the boundary condition to be $e^{\nu(10^7)}=1$. From the computed metric component $g_{tt}$ we compute the deviation function $f(r)$ defined in Eq. (\ref{eq: f func}). In the absence of matter $f(r)=1$.

In Fig. \ref{fig: metric in scaled} we plot $g_{tt}$ and $f(r)$. For all the profiles $g_{tt}$ tends to zero for values close to $r=2$. In the near region, the deviation function for NC solutions becomes almost flat, although different from unity. For the C solution $f(r)$ starts to grow for the lower value of the radius as was also found in the not scaled density profiles.

\section{Discussion and conclusion}\label{sec:Discussion and conclusion}

The current work is a step forward towards a general relativistic description of compact objects immersed in a non-trivial environment. We went beyond previous analysis where the current problem was addressed either analytically or numerically \cite{Cardoso:2020iji, Figueiredo:2023gas}. In this work for the first time, we considered non-vanishing radial pressure in the analogous manner of a compact star. We discussed both isotropic fluid configuration as well as anisotropic fluid with vanishing radial pressure. 

Using this we studied the impact of different DM distributions on the BH spacetime. We demonstrated that the metric is sensitive to the DM profile structure. However, the crucial aspect of the work is the energy condition and the sound speed. We showed that in the anisotropic case where the radial pressure vanishes, all the profiles violate the dominant energy condition near the BH. By defining the tangential sound speed we demonstrated that the tangential sound speed becomes faster than light. This feature is present for all the DM profiles considered.

\begin{figure*}
\includegraphics[width=85mm]{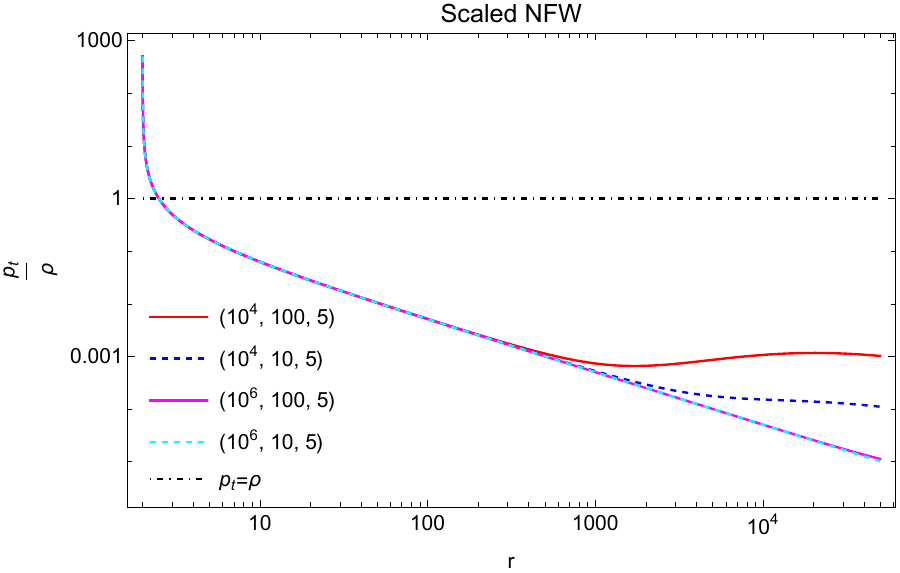}
\includegraphics[width=85mm]{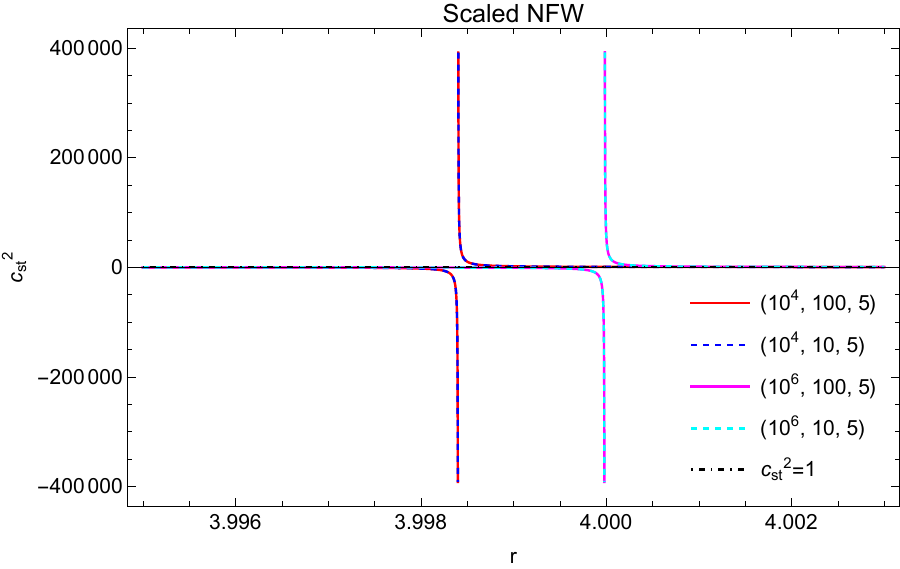}
\includegraphics[width=85mm]{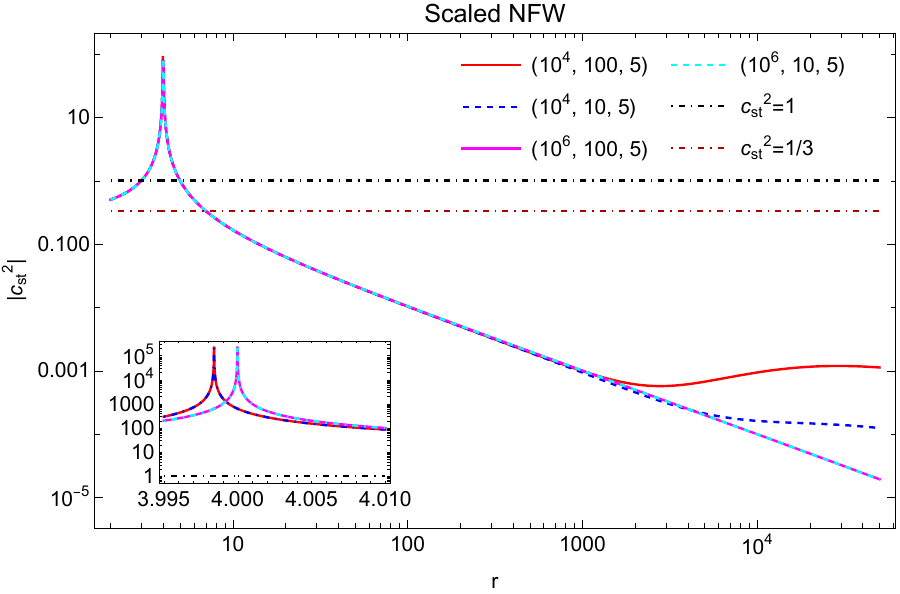}
\includegraphics[width=85mm]{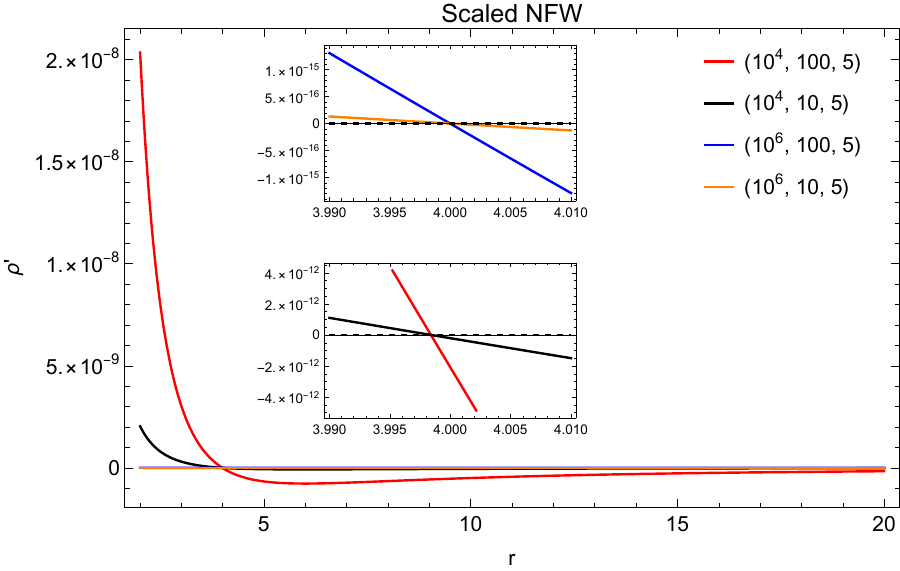}
\caption{The results for the NFW profile are shown here. In the first row, we show the behavior of the dominant energy condition and the $c_{st}^2$. At the bottom, we show $|c_{st}^2|$ and the derivative of density with respect to the radius. The connection between the vanishing $\rho'$ and the diverging sound speed is present here also. The profile is qualitatively similar to all the other profiles. Therefore, all the findings in the previous part of the current work are qualitatively valid for the NFW profile.}
\label{fig:NFW}
\end{figure*}

\begin{figure*}
\includegraphics[width=85mm]{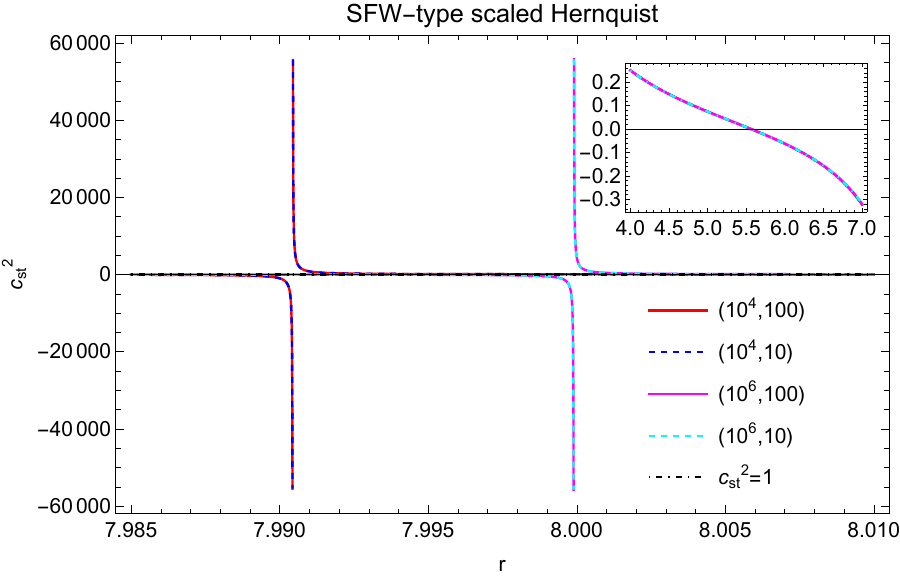}
\includegraphics[width=85mm]{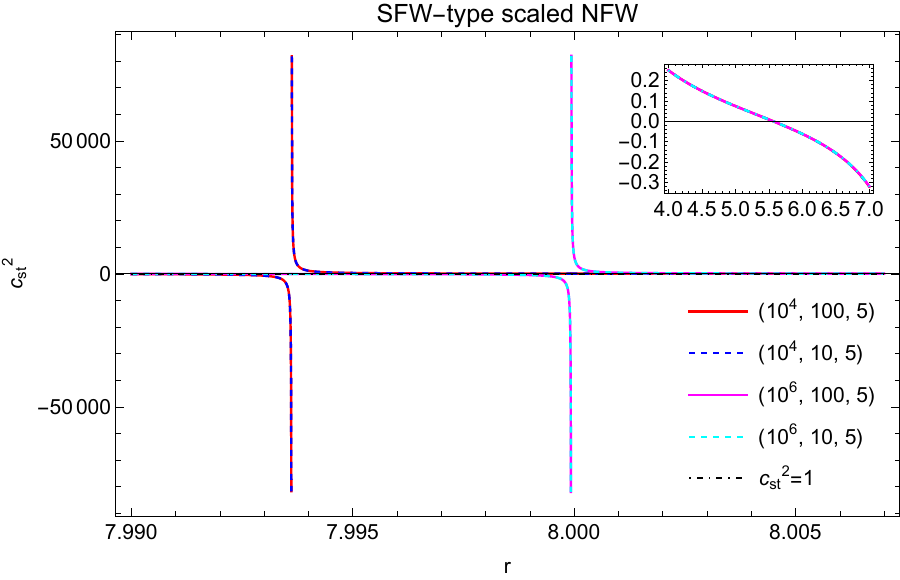}
\includegraphics[width=85mm]{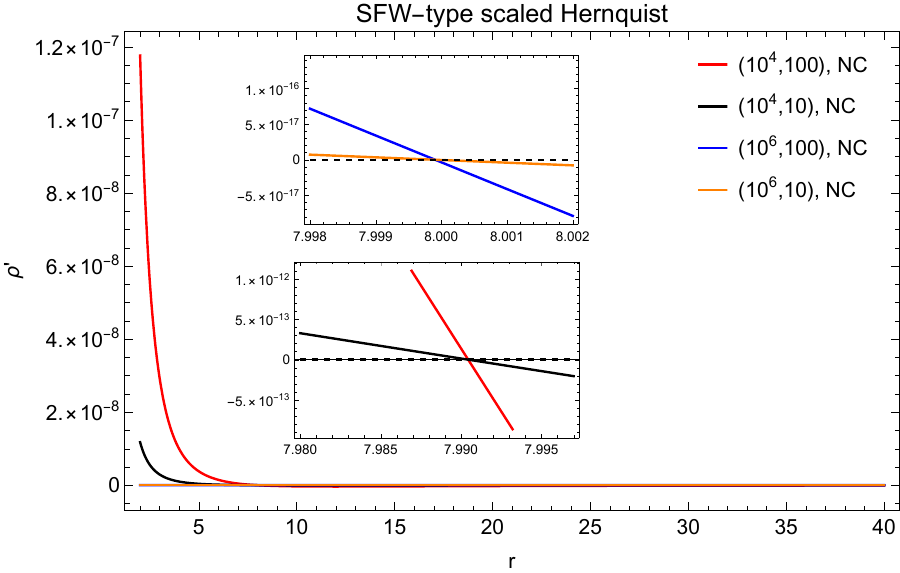}
\includegraphics[width=85mm]{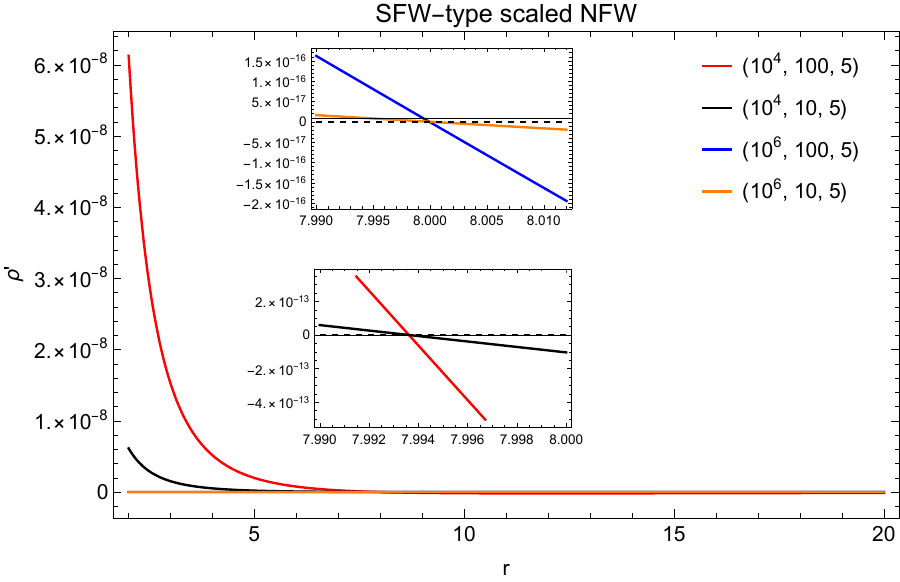}
\caption{Here the results for SFW-type scaling $(1-4/r)$ are shown. In the first row, we plot $c_{st}^2$ and we zoom in the diverging region. For all the profiles, below a certain radius $c_{st}^2$ becomes negative. In the second row, we plot the derivative of the density function with respect to the radius. We find that in most of the range of radius, the derivative is negative and in the near zone it becomes positive. The point when it becomes zero corresponds to the maxima of the density function. As a consequence $c_{st}^2$ blows up at that point and becomes negative for a smaller value of radius. Note, that the region of divergence is shifted to a larger radius for $(1-4/r)$ scaling compared to the $(1-2/r)$ scaling.}
\label{fig:scaled 4}
\end{figure*}

In the second approach, where we do not assume the radial pressure to be vanishing we find similar behavior like the anisotropic pressure. The dominant energy condition is violated near the BH. The sound speed also becomes faster than light. To make the sound speed behave physically we put up a hard condition for sound speed, such that it never goes beyond $1/3$. We solved the TOV equation under such conditions. We found that the newly found solution does not violate the dominant energy condition nor does the sound speed become faster than light. However, as a consequence, we found that the density and pressure become larger near the BH. We computed the metric with the newly found configuration and found that it is sensitive to the profile structure. The $g_{tt}$ component of the metric goes to very small values near $r\to 2$ pointing towards a possible horizon structure. We defined a deviation function that captures the deviation in $g_{tt}$ component compared to a vacuum BH. We found the deviation function shows a growth near the horizon. Despite that, the $g_{tt}$ component shows a structure that is not inconsistent with a horizon description.

In Ref. \cite{Gondolo:1999ef, Sadeghian:2013laa} it was demonstrated that the DM density near BH vanishes. Based on this, previous works had scaled the density profile to make the density vanish in the near zone. We followed the same for the latter half of our work. We took these modified density profiles. We found before imposing the sound speed condition that these profiles have more pathologies. Not only does the dominant energy condition and the tangential sound speed become superluminal but also the square of sound speeds becomes diverging and negative in the near region. By plotting the radial derivative of density we demonstrated these divergences and the negative values are connected to the maxima of the density profile. Once we put the physical sound speed condition we find the pathologies are resolved and rather than a maxima and decaying density structure the density starts to grow. This points towards the possibility that the physical sound speed under the current modeling does not allow the density to decrease or demonstrate a vanishing structure. It therefore contradicts the notion that the density should reach maxima and then vanish near BH. It remains to be seen how the possible contradiction with Ref. \cite{Gondolo:1999ef, Sadeghian:2013laa} be addressed. It can have a connection with the ``Cusp-Core" problem. It is also important to investigate in detail if the current modeling is appropriate for the DM energy-momentum tensor. It requires further detailed investigation.

Ref. \cite{Gondolo:1999ef, Sadeghian:2013laa} considered the BH to be growing adiabatically inside the DM halo, which resulted in the density maxima near the horizon. On the other hand, in Ref. \cite{Baccetti:2018otf, Terno:2019kwm} it has been shown that the regularity of $T^{\mu\nu}T_{\mu\nu}$ near horizon for a time-dependent spherically symmetric metric, i.e. $(g_{tt}, g_{rr})$, requires non-zero $T_{rt}$ component. Since the vanishing density near the horizon comes from the calculations that assume adiabatic time evolution of the BH, it is important to investigate whether modeling such a DM profile as a diagonal energy-momentum tensor is appropriate. The origin of the violation of the energy condition and the ill-behaved sound speed could also be the diagonal form of the energy-momentum tensor, at least partially. Hence, it boils down to investigating the accuracy of the assumption that the DM profile is a static structure and the resulting metric is time independent too. These aspects must be investigated in the future.

Notably, we have not assumed the solution to be a BH apriori. Along with the energy-momentum tensor model, we have only assumed the metric to be spherically symmetric and asymptotically flat. From the behavior of $g_{rr}$ we know that it diverges at $r\gtrsim 2$ since $m(r)>1$. From the solution of $g_{tt}$, we indeed find that $g_{tt}\to 0$ as $r\to 2$. However, this does not imply apriori that the points where $g_{tt}=0$ and $g^{rr}=0$ are the same point. Hence, it requires further investigations focusing on the near zone to establish if it is indeed a BH or a wormhole or shows some other structure. This needs more investigation.

\section*{Acknowledgement}
SD thanks Bhaskar Biswas for the suggestion to look into the behavior of the sound speed as well as for other discussions. SD thanks Vitor Cardoso, Sumanta Chakraborty, Prasun Dhang, and Aseem Paranjape for their discussions. SD thanks Vitor Cardoso and Andrea Maselli for reading the manuscript and providing comments.

\appendix

\section{NFW}
\label{NFW}

One density profile that we did not discuss in the main text for brevity is the Navarro-Frenk-White (NFW) profile. We scale the NFW density profile to make the density vanish near the BH. The NFW distribution can be obtained by fixing $(\alpha, \beta, \gamma) = (1, 3, 1)$ \cite{Navarro:1996gj}

\begin{equation}
    \rho_{NFW, S} = \left(1-\frac{2}{r} \right)  \rho_0\left(\frac{r}{a_0}\right)^{-1}\left[1 + \left(\frac{r}{a_0}\right) \right]^{-2}
\end{equation}

The NFW model is well-known for predicting a mass function that diverges logarithmically as $r$ approaches infinity. To address this divergence, we introduce a radial cut-off $r_c$ such that $M_{\text{Halo}}(r > r_c) = 0$ similar to that of Ref. \cite{Figueiredo:2023gas}. We set it to be $r_c=5a_0$. With the density at hand, we plot $p_t/\rho$, tangential sound speed as well as $d\rho_{NFW, S}/dr$ in Fig. \ref{fig:NFW}. We label individual profile by $(a_0,M_{\rm Scale},r_c/a_0)$. It can be seen from the plots, that the NFW profile also demonstrates the previously discussed pathologies. For brevity, we are not showing NC or C solutions of the corresponding TOV equations. We found them to demonstrate features similar to the other profiles.

\section{Sadeghian-Ferrer-Will (SFW) type profile}
\label{app: scale 4}

In the main text along with the actual density profiles, we also showed the results of scaled density profiles. However, the scaling was done in a manner so that the density vanishes at the horizon, i.e. $r=2$. Ref. \cite{Sadeghian:2013laa} studied what will be the density profile of a halo if a black hole is grown adiabatically within a preexisting halo. They found the density to vanish rather at $r=4$. The disappearance of particle density at $r=4$ is understood in terms of stable circular orbits. This value corresponds to the radius of the unstable circular orbit within the Schwarzschild geometry for a particle on the verge of escaping, characterized by energy per unit mass $E = 1$ and having an angular momentum per unit mass $L = 4$. 
A particle with $L \geq 4$ and $E \leq 1$ has an inner turning point at $r \geq 4$. 
Consequently, any particle managing to approach $r = 4$ inevitably falls into the black hole. Therefore, the particle density vanishes at $r \leq 4$. For further details see \cite{Sadeghian:2013laa}.

For this reason, here we show the result for such kind of vanishing density. Here we take $\rho_{Hern/NFW} \to (1-4/r)\rho_{Hern/NFW}$. From the study in the main text, we showed that the vanishing of density implies there is a maximum in the density and the density decreases afterward. We demonstrated that this is the region where the sound speed behaves unphysically, precisely because of the density structure. Therefore having a different scaling will not change the qualitative features. The region of divergences will just get shifted.

In the first row of Fig. \ref{fig:scaled 4} we plot $c_{st}^2$ and we zoom in on the diverging region. For all the profiles, below a certain radius, $c_{st}^2$ becomes negative even for the current scaling. In the second row, we plot the derivative of the density function with respect to the radius. We find that in most of the range of radius, the derivative is negative and in the near zone it becomes positive. The point when it becomes zero corresponds to the maxima of the density function. As a consequence $c_{st}^2$ blows up at that point and becomes negative for a smaller value of radius. These features are exactly similar qualitatively to the $(1-2/r)$ scaling we discussed in the main text. In the current scaling, the density vanishes everywhere $r<4$, unlike the previous case where it vanishes at the horizon. As a result, in the current case, the region of divergence is shifted to a larger radius compared to the $(1-2/r)$ scaling. This bolsters the point that the maxima of density is the reason behind the unphysical behavior of sound speed. 

Interestingly for this scaling, in the region with the radius smaller than that of the divergence points, there exists a patch where sound speed can be physical. In the subplot of Fig. \ref{fig:scaled 4} this region is shown. Below the divergence point, the sound speed becomes negative similar to the other scaling. However, later it starts to grow and around $r\sim 5.5$, it reaches zero. From $\sim 5.5$ to $r=4$ it becomes positive and subluminal. We are not showing the region below $r=4$, as the density vanishes in that region. This result implies that depending on the scaling, below the divergence points physical sound speed can exist in certain patches. However, it can not resolve the divergences and the negative sound speed. Therefore the maxima of the density is bound to create these pathologies. However, depending on how and where the density reaches maxima, physical sound speed patches can form near the BHs.

\bibliography{references}

\end{document}